\documentclass[a4paper,11pt]{article}
\pdfoutput=1 

\usepackage{jcappub} 

\usepackage[T1]{fontenc} 
\usepackage{adjustbox}

\title{\boldmath   Bondi-Hoyle-Lyttleton accretion onto a rotating black hole with
  ultralight scalar hair}
\author[a,1]{Alejandro Cruz-Osorio,\note{Corresponding author.}}
\author[a,b,c]{Luciano Rezzolla} 
\author[d]{Fabio D. Lora-Clavijo}
 \author[e,f]{Jos\'e Antonio Font}
\author[g]{and Carlos Herdeiro}
\author[g]{Eugen  Radu} 

\affiliation[a]{Institut f{\"u}r Theoretische Physik, Goethe Universit\"at, Max-von-Laue-Stra{\ss}e 1, 60438 Frankfurt, Germany}
\affiliation[b]{Frankfurt Institute for Advanced Studies, Ruth-Moufang-Strasse 1, 60438 Frankfurt, Germany}
\affiliation[c]{School of Mathematics, Trinity College, Dublin 2, Ireland}
\affiliation[d]{Grupo de Investigaci\'on en Relatividad y Gravitaci\'on, Escuela de F\'isica, Universidad Industrial de Santander A. A. 678, Bucaramanga 680002, Colombia}
\affiliation[e]{Departamento de Astronom\'ia y Astrof\'isica, Universitat de Val\`encia, Dr. Moliner 50, 46100, Burjassot (Val\`encia), Spain}
\affiliation[f]{Observatori Astron\`omic, Universitat de Val\`encia, C/ Catedr\'atico Jos\'e Beltr\'an 2, 46980, Paterna (Val\`encia), Spain}
\affiliation[g]{Departamento de Matem\'atica da Universidade de Aveiro and CIDMA, Campus de Santiago, 3810-183 Aveiro, Portugal}

\emailAdd{osorio@itp.uni-frankfurt.de}

\abstract{
We present a numerical study of relativistic Bondi-Hoyle-Lyttleton (BHL)
accretion onto an asymptotically flat black hole with synchronized
hair. The hair is sourced by an ultralight, complex scalar field,
minimally coupled to Einstein's gravity. Our simulations consider a
supersonic flow parametrized by the asymptotic values of the fluid
quantities and a sample of hairy black holes with different masses,
angular momenta, and amount of scalar hair. For all models, steady-state
BHL accretion solutions are attained that are characterized by the
presence of a shock-cone and a stagnation point downstream. For the
models of the sample with the largest component of scalar field, the
shock-cone envelops fully the black hole, transitioning into a bow-shock,
and the stagnation points move further away downstream. Analytical
  expressions for the mass accretion rates are obtained after fitting the
  numerical results, which can be used to analyze black-hole formation
scenarios in the presence of ultralight scalar fields. The formation of a
shock-cone leads to regions where sound waves can be trapped and resonant
oscillations excited. We measure the frequencies of such quasi-periodic
oscillations and point out a possible association with quasi-periodic
oscillations in the X-ray light curve of Sgr~A* and microquasars.
} 

\begin{document}
\maketitle
\flushbottom

\section{Introduction}
\label{sec:intro}
Active galactic nuclei (AGNs) are powered by the accretion of gas onto
their central supermassive black holes (SMBHs) and emit intense
electromagnetic radiation in a broad range of frequencies. Observations
of the plasma moving around regions close to the horizon of SMBHs are now
possible thanks to the Event Horizon Telescope (EHT), which recently
imaged the black hole at the center of the galaxy
M87~\cite{EHT_M87_PaperI} and of the Milky
Way~\cite{EHT_SgrA_PaperI}. Black holes are one of the most exciting
outcomes of relativistic astrophysics and general relativity. From a
mathematical point of view, there are no restrictions on the black hole's
mass. On the other hand, astronomical observations clearly indicate the
evidence for astrophysical black holes in two mass windows only: the
stellar-mass range ($4\, M_{\odot} \lesssim M_{\rm BH} \lesssim 80\,
M_{\odot}$)~\cite{Remillard2006, Abbot2016-GW-detection-prl} and the
supermassive range ($10^{6}\, M_{\odot} \lesssim M_{\rm BH} \lesssim
10^{9}\, M_{\odot}$)~\cite{Ferrarese2005, Kormendy2013,
  EHT_M87_PaperI}. However, the LIGO-Virgo-KAGRA scientific collaboration
(LVK) has reported the detection of gravitational-wave transient
GW190521\cite{Abbott2020c}, an intermediate-mass black hole of $150\,
M_{\odot}$ resulting from the merger of two black holes of masses $85 \,
M_{\odot}$ and $66 \, M_{\odot}$. The nature of this source remains
somewhat uncertain (see e.g.,~\cite{Bustillo2021, Cruz2021a,
  DeLuca2021,Shibata2021b, Gamba2022} for alternative
interpretations). In addition, explaining the masses of SMBHs is still an
important open problem in astrophysics. One approach to this problem
assumes that the large masses of SMBHs are the result of matter accretion
onto black hole seeds~\cite{Lightman1977, Begelman2006,
  Volonteri2010}. Therefore, understanding the details of matter
accretion onto black holes is of great relevance on a series of timely
problem in gravitational-wave astronomy.

When considering the dynamics of gas around a point mass, one of the most
used models was developed by Bondi, Hoyle and Lyttleton
(BHL)~\cite{Hoyle1939, Bondi1944, Bondi52}. The BHL model assumes
accretion onto a central compact object moving in a {homogeneous
  distribution of gas} and has been widely {employed} in different
astrophysical scenarios, via numerical studies in the Newtonian regime
\cite{Hunt1971, Matsuda1987, Fryxell1988, Sawada1989, Livio1991, Ruffert1994,
  Ruffert1994b, Ruffert1997, Foglizzo1999, Foglizzo2005, Mellah:2015sja,
  Beckmann:2018xfi}. A historical {overview} of the development and
numerical modeling of {BHL accretion studies} can be found in
\cite{Edgar2004, Rezzolla_book:2013}. {The validity of a Newtonian
  approach breaks down when the accreting object is a black hole or a
  neutron star, as general relativistic effects become important,
  especially when capturing the flow dynamics close to the black hole
  horizon (see, e.g.,~Ref.~\cite{Cruz2020b}, where a Newtonian and
  relativistic treatment of mass accretion in a common-envelope scenario
  are contrasted). This motivated the first numerical works dedicated to
  studying the morphological patterns in the vicinity of a black hole as
  well as to compute the associated accretion rates of mass and
  momentum~\cite{Petrich89, Font98a, Font98c, Font98d,
    Font1999b}. Subsequently, many simulations have investigated the BHL
  mechanism in the context of relativistic astrophysics~\cite{Donmez2010,
    Cruz2012, Lora2013, Cruz2013}, becoming increasingly realistic (and
  more complex) by including magnetic fields~\cite{Penner2011, Kaaz2022, Gracia-Linares2023},
  radiative terms~\cite{Zanotti2011}, density and velocity
  gradients~\cite{Lora2015219,Cruz2016}, and small rigid bodies around
  the black hole~\cite{Cruz2017}. More recently, the role of BHL
  accretion in the common envelope phase in the evolution of binary
  systems has also been investigated by~\cite{Cruz2020b}.}

If SMBHs result from accretion onto black-hole seeds, a key question is
to determine the interplay between baryons and dark matter in this
{scenario~\cite{Lightman1977, Read2003, King2006, Guzman2011a,
    Guzman2011b, Lora2014}.}  In this respect, scalar fields have been
proposed to play the role of dark matter, first at galactic scale
\cite{Matos1998vk, Matos2000} and then at cosmic scale
\cite{Sahni:1999qe, Matos:2000ng}. Scalar fields interacting with gravity
{could therefore play a role} in understanding the physics associated
with black holes, in particular {its potential impact on} SMBH
formation.

Scalar field accretion onto black holes has been already explored in
different works. In the context of the ultra-light scalar fields proposed
as constituents of dark matter halos, it has been shown that the
accretion rate would be small over the lifetime of a typical halo, and
hence {the central SMBHs could coexist with scalar-field
  halos~\cite{UrenaLopez:2002du}.} {The study of scalar-field accretion
  has also been carried out on the spacetime background of boson
  stars~\cite{LoraClavijo:2010xc} and black
  holes~\cite{CruzOsorio:2010qs, Cruz2010, Cruz2010a} using ``Scri''
  ($\mathcal{I}$)-fixing conformal compactification, that is, including
  future null infinity $\mathcal{I}^+$ in the numerical simulations. In
  the former case, the scalar field shows quasi-normal mode oscillations
  only for boson star configurations that are compact enough, where no
  signs of tail decay were found. In the latter, an extremely high
  dispersion rate of the scalar field density was found.}

The possible existence of equilibrium configurations between ultralight
scalar fields and black holes in general relativity has also been
addressed in the literature. In~\cite{Barranco2011, Barranco2012} it was
shown that a massive scalar field surrounding a Schwarzschild black hole
can survive for arbitrarily long times, as quasi-bound states, forming
scalar ``wigs''. Moreover, in the context of the nonlinear evolution of
massless scalar fields (also accounting for the evolution of the
spacetime), Ref.~\cite{Guzman:2012jc} showed that scalar fields are
allowed to survive outside black holes and may eventually have lifetimes
consistent with cosmological timescales. Further studies have confirmed
the existence of such long-lived states~\cite{Witek2013, Sanchis2015,
  Barranco2017, Aguilar2022}. In addition to the idea that scalar fields
can survive for long timescales around black holes,
Ref.~\cite{Degollado2014} showed that the gravitational radiation emitted
by a Schwarzschild black hole when a massive scalar field is accreted
exhibits a distinctive signature, a wiggly tail, in the late-time
behavior of the signal.

Eventually, true equilibrium states between ultralight scalar fields and
black holes in general relativity were found~\cite{Herdeiro2014,
  Herdeiro2015}. A new family of solutions of the Einstein-Klein-Gordon
field equations, where the scalar field is complex, massive, and
minimally coupled to general {relativity, was} derived in
Ref.~\cite{Herdeiro2014}, to which we will hereafter refer to as ``hairy
black holes'' (HBHs). These solutions describe asymptotically flat
rotating black holes with scalar hair that are regular outside of the
event horizon and have the following properties:
\begin{itemize}
\item are supported by rotation and have no static limit
\item have a conserved continuous Noether charge $q$ measuring the
  scalar hair
\item have a Kerr limit when $q=0$
\item have a solitonic limit when $q=1$
\end{itemize}
For more details on the construction of the solutions and their different
properties see~\cite{Herdeiro2015}. The stability of {these} hairy Kerr
black holes was discussed in~\cite{Ganchev2018, Degollado2018}. In
particular, it was shown through numerical analysis that the superradiant
instability appears for those solutions~\cite{Ganchev2018}, but the
corresponding timescale increases with the black-hole mass and can be
larger than the age of the Universe for sufficiently (super)massive black
holes~\cite{Degollado2018}.  Therefore, these HBHs provide an interesting
model to study the interplay between the accretion of baryonic matter and
dark matter in the context of SMBH-formation modeling.

There exist studies aimed at assessing the astrophysical relevance of the
aforementioned HBHs~\cite{Herdeiro2014, Herdeiro2015}. For instance,
Ref.~\cite{Cunha2015} investigated the shadow produced by HBHs using
backward ray tracing; see also~\cite{Cunha:2016, Cunha:2019ikd}. The
extension to electrically charged black holes (Kerr-Newman solution) with
scalar hair can be found in~\cite{Delgado2016}. Properties of geodesic
motion and its relation to quasi-periodic oscillations (QPOs) have been
considered in~\cite{Franchini2016}, while the X-ray spectroscopy has been
addressed in~\cite{Ni2016a}. More recently, HBHs have been used to build
relativistic stationary models of magnetized tori in the test-fluid
limit~\cite{Gimeno-Soler:2019, Gimeno-Soler2021}.

In this paper, we initiate a program to numerically model relativistic BHL
accretion onto an asymptotically flat rotating
HBH~\cite{Herdeiro2014}. In this first investigation, we assume the
test-fluid approximation and neglect the evolution of the gravitational
field and of the scalar field, concentrating only on the dynamics of the
fluid. The simulations have revealed that the morphology of the flow past
HBHs is similar to that found for BHL accretion onto a Kerr black hole
and that also in this case an upstream bow-shock and a downstream
shock-cone can be produced and whose properties depend on the relative
strength of the scalar field over the central black hole. At the same
time, the phenomenology of the flow is also richer because of the
increased possibility of trapping pressure modes in the cavities that are
produced, either by the fluid or by the scalar field, and that can in
principle be observed in terms of QPOs.

The organization of the paper is the following: in Sec. \ref{sec:HBH} we
briefly describe the theory associated with our specific family of HBHs
as well as the sample of (four) configurations we consider. In Sec.
\ref{sec:RH} we present the mathematical framework for the
general-relativistic hydrodynamics equations and the numerical setup
used. Section \ref{sec:results} discusses the accretion dynamics and
morphology, along with the mass and angular momentum accretion rates for
all HBHs models of our study. Moreover, in Sec. \ref{sec:AA} we show an
astrophysical application of our model by measuring frequencies of
oscillation of resonant cavities of our system and comparing them with
the QPO frequencies observed in Sgr~A* and several microquasar
sources. Finally, Sec.~\ref{sec:discussion} summarizes our
conclusions. Throughout the paper, Greek indices run from 0 to 3 while
Latin ones run from 1 to 3. Additionally, equations are written in
geometrized units where $G=c=1$.

\section{Rotating Black Holes with Scalar Hair}
\label{sec:HBH}

\begin{table*} 
\begin{center}
  \renewcommand{\arraystretch}{1.4}
  \setlength\tabcolsep{3.pt}
  \begin{tabular}{lcc|cc|cc|cc|ccc}
\hline
 Model  &$M_{\rm ADM}$&$J_{\rm ADM}$&$a_{\rm ADM}$& $a_{\rm BH}$ & $p_{\rm EH}$&$q_{\rm EH}$&$p$&$q$&$\Omega_{\rm EH}$& $r_{\rm EH}$& $M_{\phi}/M_{\rm BH}$\\
\hline
\hline
\texttt{HBH-a} &$0.415$  & $0.172$ & $0.999$  & $0.971$ & $0.947$ & $0.872$ & $0.053$ & $0.128$ & $0.975$ & $0.200$ & $0.06$ \\
\texttt{HBH-b} &$0.933$  & $0.739$ & $0.849$  & $2.075$ & $0.251$ & $0.154$ & $0.749$ & $0.846$ & $0.820$ & $0.100$ & $2.98$ \\
\texttt{HBH-c} &$0.539$  & $0.495$ & $1.704$  & $0.617$ & $0.091$ & $0.003$ & $0.909$ & $0.997$ & $0.980$ & $0.101$ & $9.99$ \\
\texttt{HBH-d} &$0.975$  & $0.850$ & $0.894$  & $5.519$ & $0.018$ & $0.002$ & $0.982$ & $0.998$ & $0.680$ & $0.040$ & $54.56$ \\
\hline
\end{tabular}
\caption{Spacetime properties corresponding to our HBH solutions:
  reported are the ADM mass $M_{\rm ADM}$ and total angular momentum and
  $J_{\rm ADM}$, together with the fractions of energy $(p$) and angular
  momentum ($q$) measured at the event horizon (subscript ${\rm EH}$) and
  in the exterior scalar field. Thus $p_{\rm EH}:= M_{\rm BH}/M_{\rm
    ADM}$, $q_{\rm EH} := J_{\rm BH}/J$, $p:= M_{\phi}/M_{\rm
    ADM}=1-p_{\rm EH}$, and $q:= J_{\phi}/J_{\rm ADM}=1-q_{\rm EH}$. For
  convenience we redefine the dimensionless spin parameters as $a_{\rm
    ADM}:=J_{\rm ADM}/M_{\rm ADM}^2$ and $a_{\rm BH}:=J_{\rm BH}/M_{\rm
    BH}^2$. The black-hole mass and angular momentum, $M_{\rm BH}$ and
  $J_{\rm BH}$, are computed as Komar surface integrals; the scalar field
  mass and angular momentum, $M_{\phi}$ and $J_{\phi}$, are computed as
  volume integrals over the corresponding Komar densities. The last three
  columns provide the angular velocity and the event-horizon radial
  coordinate of the HBH, $\Omega_{\rm EH}$, $r_{\rm EH}$, as well as the
  mass ratio between scalar field and black hole, which we will take as
  an ordering parameter to measure the relative strength of the
  scalar-field hair. }
\label{tab:1}
\end{center}
\end{table*}

{A HBH} can be envisaged as a BH whose horizon is surrounded by, and
in synchronous rotation with, a scalar field cloud that decays
exponentially fast. The geometry and scalar field profile of {HBHs}
read~\cite{Herdeiro2014} 
\begin{eqnarray}
ds^2 &=&  -  e^{2{\cal F}_{0}} N dt^{2} + e^{2{\cal F}_{1}} \left(  \frac{dr^{2}}{N} + r^{2}d\theta^{2}\right) \nonumber  \\
&&+  e^{2{\cal F}_{2}} r^{2} \sin ^{2} \theta \left( d\varphi - {\cal W} dt\right)^{2} \,,
\label{eq:metric}
\end{eqnarray}
where
\begin{eqnarray}
N    &:=&  1- \frac{r_{\rm EH}}{r } \,,  \\
\Psi &:=& \phi(r,\theta) e^{i(m\varphi-\omega t)} \,.
\label{eq:scalarfield}
\end{eqnarray}
Here, $r_{\rm EH}$ marks the radial location of the event horizon,
$\omega$ is the scalar field frequency, which is assumed to be positive,
$m = \pm 1, \pm 2, ...$ is the azimuthal harmonic index, which for the
solutions herein is always $m=1$, ${\cal F}_{0}, {\cal F}_{1}, {\cal
  F}_{2}, {\cal W}$ and $\phi(r,\theta)$ are unknown functions of $(r,
\theta)$, which satisfy the asymptotically flat spacetime condition
\begin{eqnarray}
\lim\limits_{r \to \infty} {\cal F}_{i}=1 \ , \qquad \lim\limits_{r \to
  \infty} {\cal W}=0 \ , \qquad \lim\limits_{r \to \infty} {\cal
  \phi}=0\,.
\end{eqnarray}

\begin{figure*}
\centering
\includegraphics[width=1.0\textwidth]{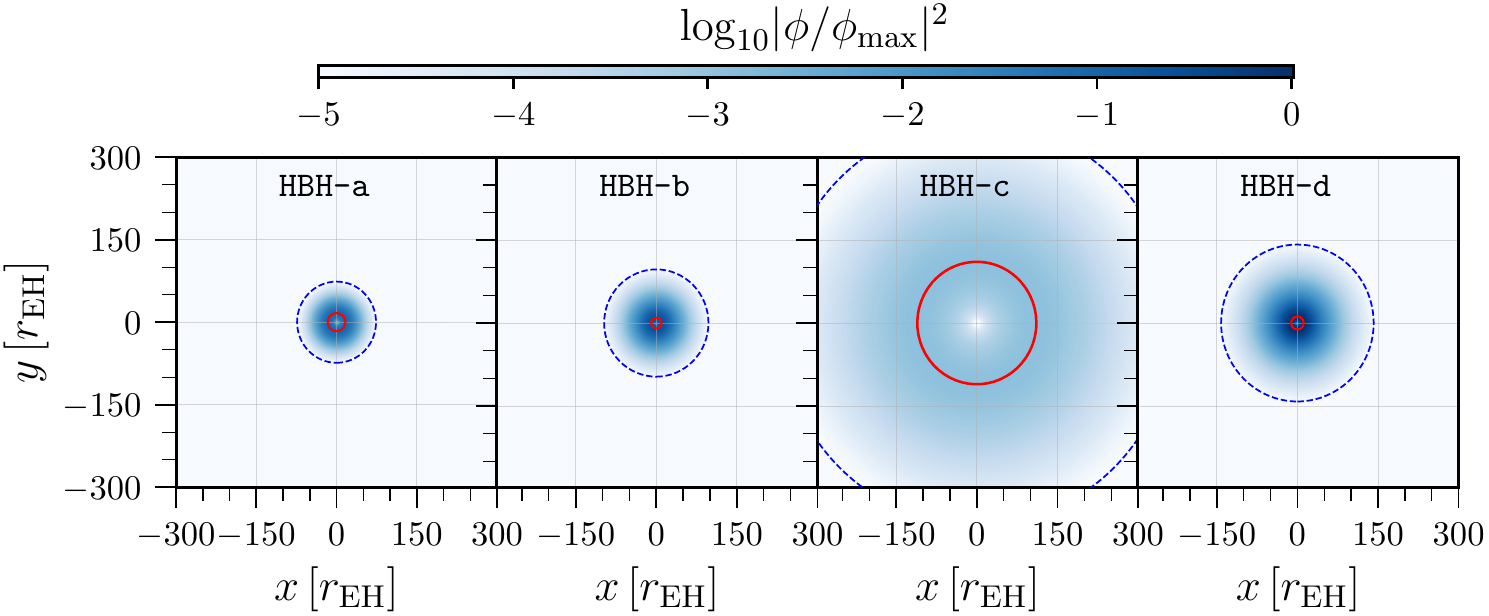}
\caption{Two-dimensional distribution in the equatorial plane of the
  ultralight scalar hair density around the black hole (the
  three-dimensional morphology is that of a torus of scalar hair).  The
  red circles show the location of the maximum amplitude, $\phi_{\rm
    max}$.} \label{fig:hbhsolution}
\end{figure*}

\begin{figure}
\centering
\includegraphics[width=0.5\textwidth]{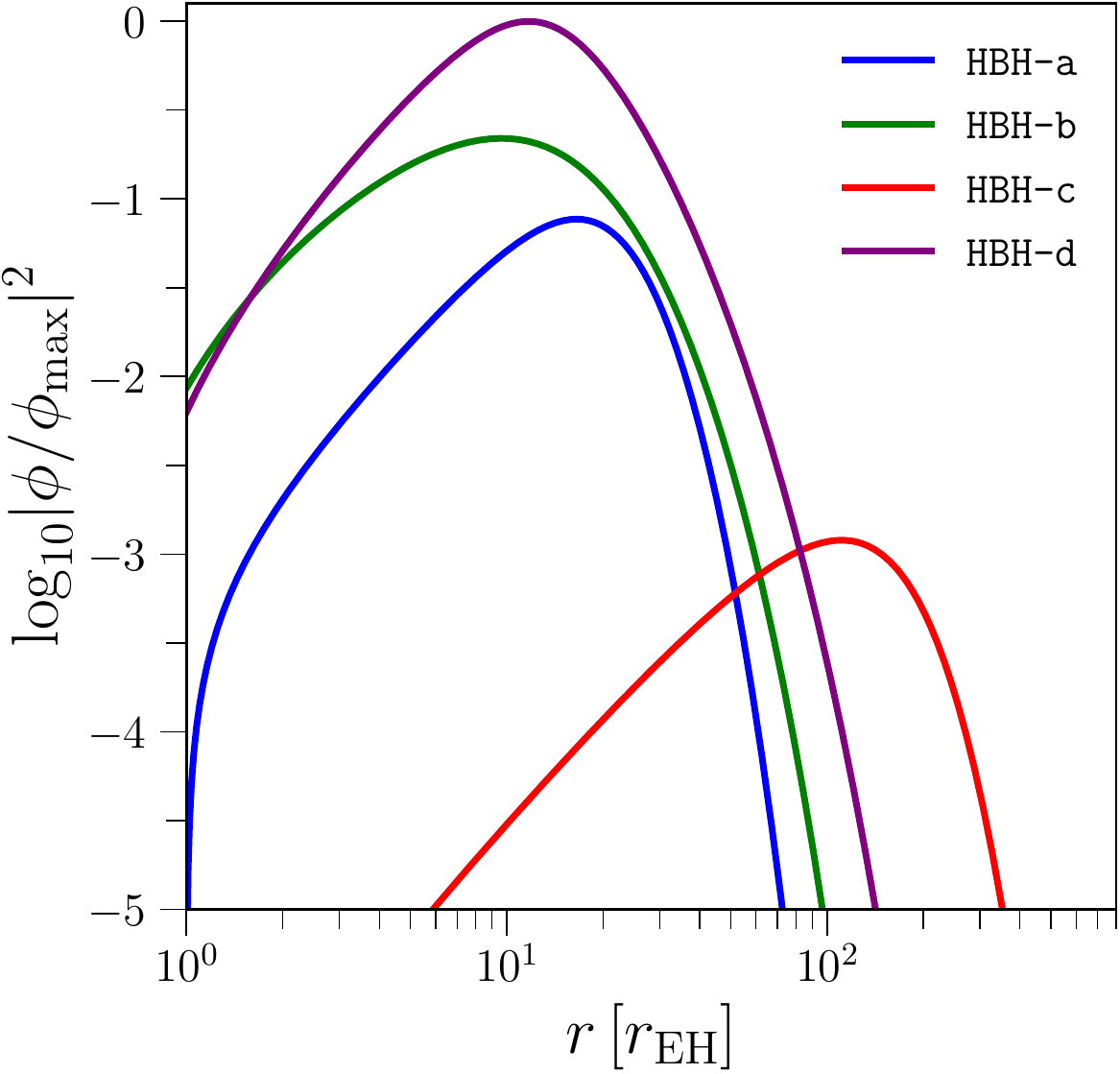}
\caption{One-dimensional radial profiles of normalised scalar field
  amplitude, where $\phi_{\rm max}=3.2\times 10^{-1}$ is the maximum
  amplitude {for} model \texttt{HBH-d}. } \label{fig:hbhsolution_1}
\end{figure}

The complex scalar field $\Psi=\Psi^{R} + i \Psi ^{I}$, where $\Psi^{R}$
is its real part and $\Psi^{I} $ the imaginary one, respectively, is
assumed to be minimally coupled to gravity. In this way, the
Einstein--Klein-Gordon (EKG) evolution equations can be obtained by
performing the variation of the action with respect to the metric and the
complex scalar field. The action ${\cal S}$ and the energy-momentum
tensor $T_{\sigma \lambda}$ of this system can be written as follows
\begin{eqnarray}
{\cal S}&=&\int d^{4}x \sqrt{-g} \left[ \frac{R}{16\pi} -
  \frac{1}{2}g^{\sigma \lambda} ( \partial_{\sigma}\Psi^{*} \partial_{\lambda}\Psi +
  \partial_{\sigma}\Psi \partial_{\lambda}\Psi^{*} ) \right. \nonumber \\
  && \left. -
  \mu^{2} \Psi^{*} \Psi \right] \,,\label{eq:action}\\ \nonumber
\\ T_{\sigma \lambda} &:=& \partial_{\sigma}\Psi^{*} \partial_{\lambda}\Psi +
\partial_{\sigma}\Psi \partial_{\lambda}\Psi^{*} \nonumber \\
&-& g_{\sigma \lambda}
\left[ \frac{1}{2}g^{\gamma\delta}\left( \partial_{\gamma}\Psi^{*}
  \partial_{\delta}\Psi + \partial_{\gamma}\Psi \partial_{\delta}\Psi^{*} \right) + \mu^{2}
  \Psi^{*} \Psi \right]\,, \label{eq:emtensor}
\end{eqnarray}
where $\Psi^{*}$ is the complex conjugate of $\Psi$, $g$ is the
determinant associated with the components of the 4-metric $g_{\sigma
  \lambda}$, $R$ is the Ricci scalar, and $\mu$ is the mass of the scalar
field.

The HBH solutions of the EKG equations employed in our numerical
simulations are the same as in~\cite{Herdeiro2014,
  Herdeiro2015,Cunha2015}. Specifically, we consider four illustrative
solutions for hairy {black holes}. The corresponding scalar field masses
and angular momenta are in the range $M_{\phi} \in [0.022 , 0.957]$ and
$J_{\phi} \in [0.022, 0.848]$, where the physical quantities are
displayed in units of the scalar field mass $\mu$. Some relevant physical
parameters of the HBHs used in our simulations are summarized in
Tab.~\ref{tab:1}. As it can be seen from this table, the relative
contribution of the scalar field to the mass and angular momentum is much
higher for the last solutions (\texttt{HBH-c,d}), which are therefore
\textit{hairier}, whereas the first ones (\texttt{HBH-a,b}) are closer to
Kerr. It is worth mentioning that in this work we focus on the
  impact of changing $M_{\phi}/M_{\rm BH}$ rather than studying the other
  degrees of freedom of the problem, such as the spin of the black hole
  or the adiabatic index of the gas. The impact of these different
  parameters will be the subject of future studies.

The numerical domain employed to construct the
  HBH solutions is $r\in [1.1\,r_{\rm EH}, 200\,M_{\rm BH}]$ and $\theta
  \in [0, \pi]$ (see also Sec.~\ref{sec:results} for a description of the
  hydrodynamical grid). Additionally, since the numerical simulations of
BHL accretion will be performed in the equatorial plane, $\theta=\pi/2$,
we specialize the HBH solutions to this
plane. Figure~\ref{fig:hbhsolution} shows the two-dimensional morphology
of the normalized scalar field around the black hole (left panels, in
Cartesian coordinates), where the red circle marks the location of {the
  maximum amplitude of the scalar field. Correspondingly,
  Fig.~\ref{fig:hbhsolution_1} reports the radial distribution of the
  normalized scalar field amplitude.} In each solution, the scalar field
has a different mass, angular velocity, strength, and distribution in
order to have a self-consistent solution, satisfying the constraints of
the EKG system.

\section{Problem setup}
\label{sec:RH}

{The general} relativistic hydrodynamics equations in the $3+1$
spacetime decomposition are written in a conservative form following the
Valencia formulation~\cite{Banyuls97}
\begin{align}
\partial_{t} (\sqrt{\tilde{\gamma}} \, \boldsymbol{U}) + \partial_{i}
(\sqrt{\tilde{\gamma}} \, \boldsymbol{F}^{i}) = \sqrt{\tilde{\gamma}} \,
\boldsymbol{S} \,, \label{eq:conservationlaw}
\end{align}
where for the equatorial case $i,j = r, \phi$, and the conserved
variables $\boldsymbol{U}$ and fluxes $\boldsymbol{F}^{i}$ are
respectively
\begin{align}
\boldsymbol{U} = \left[ 
\begin{array}{c}
D  \\
S_{j}  \\
\tau
\end{array}\right] 
\,, \ \qquad 
\boldsymbol{F}^{i} = \left[
\begin{array}{c}
\mathcal{V}^{i} D \\
\alpha W^{i}_{j} - \beta^{i} S_{j} \\
\alpha (S^{i}-v^{i} D) - \beta^{i} \tau \end{array}\right] \,, \label{eq:uandflux}
\end{align}
where $\mathcal{V}^{i} := \alpha v^{i} - \beta^{i}$ are the
components of the transport velocity, and $W^{ij} := S^{i} v^{j}
+p\tilde{\gamma}^{ij} $ are the spatial components of the stress-energy
tensor, $\alpha$, {$\tilde{\gamma}$}, and $\beta^{i}$ are the lapse
function, the determinant of the 3-metric, and {the shift vector
  components, respectively.}  The conserved quantities in the Eulerian
frame are given by the following relations
\begin{eqnarray}
D &:=& \Gamma \rho, \\
S_j &:=& \rho h \Gamma^2 v_j, \\
\tau &:=& \rho h \Gamma^{2} - p - D,
\end{eqnarray}
where $\Gamma := \alpha u^{0}$ is the Lorentz factor with respect to the
Eulerian observer, $\rho$ is the rest-mass density, $h=1+ \epsilon +
p/\rho$ is the specific {enthalpy}, $p$ is the fluid pressure and
$\epsilon$ the specific internal energy. It is worth mentioning that the
fluid corresponds to a perfect one {which we model with an ideal gas
  equation of state.} Finally, the sources $\boldsymbol{S}$ in the
right-hand-side of equation \eqref{eq:conservationlaw} are
\begin{align}
\boldsymbol{S} = 
\left[
\begin{array}{c}
0  \\
\frac{1}{2}\alpha W^{ik}\partial_{j}\tilde{\gamma}_{ik} + S_{i}\partial_{j}\beta^{i} - U\partial_{j}\alpha \\
\frac{1}{2} W^{ik} \beta^{j} \partial_{j} \tilde{\gamma}_{ik} + W_{i}^{j}\partial_{j}\beta^{i} - S^{j} \partial_{j} \alpha
\end{array}
\right] \,. \label{eq:source}
\end{align}

A well-known feature of the above equations is that after every time step
it is mandatory to compute the primitive variables ($\rho,\, \Gamma
v^{i},\,p$) in order to compute the numerical fluxes, through some
implicit procedure. In this work, we are using the method ``1DW" to
recover the Lorentz factor in order to avoid superluminal velocities, as
prescribed in the \texttt{BHAC} code~\cite{Porth2017, Olivares2018a,
  Olivares2019}.

\begin{figure*}
\centering
\includegraphics[width=1.0\textwidth]{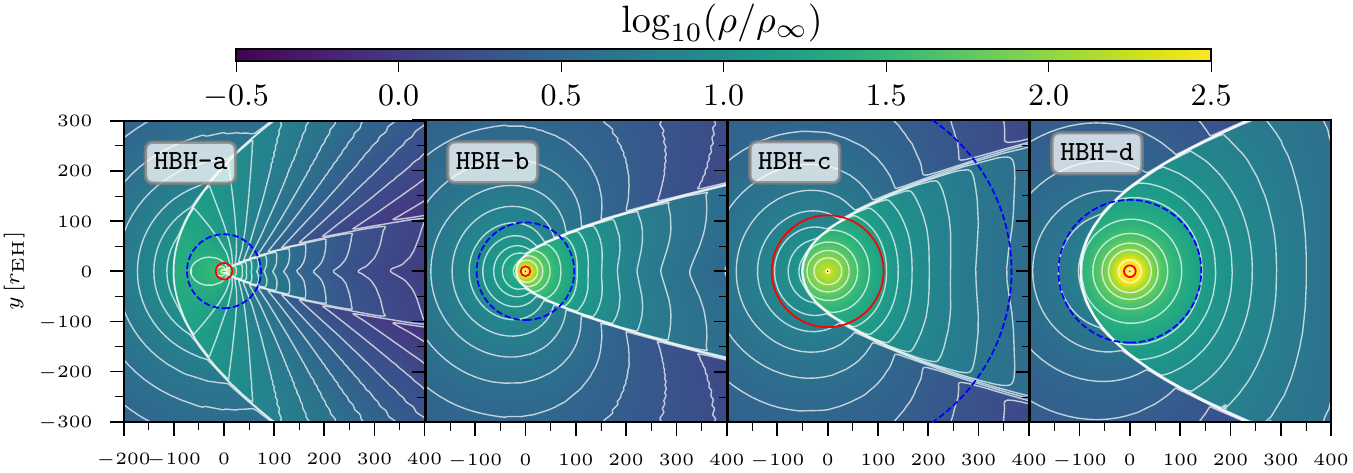}
\includegraphics[width=1.0\textwidth]{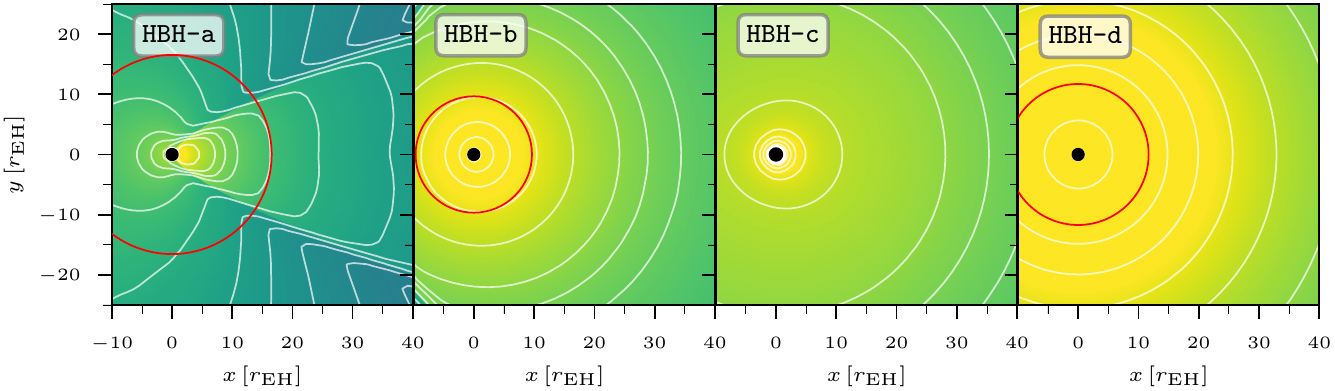}
\caption{Two-dimensional distributions (colormap) and isocontours (grey
  lines) of the rest-mass density normalized by its initial value
  $\rho_{\infty}$ and at the end of the numerical evolution
  $t=10000\,M_{\rm BH}$ for the four HBHs considered here. Red solid
  circles mark the location of the maximum scalar field $\phi_{\rm max}$,
  while the blue dashed circles mark the outer edge of the torus of
  scalar hair so that the two circles provide a view of the torus
  size. The top and bottom rows report respectively large- and
  small-scales views of portions of the computational domain so that the
  different parts of the dynamics can be appreciated.} \label{fig:2Drho}
\end{figure*}

\begin{figure}
\centering
\includegraphics[width=0.5\textwidth]{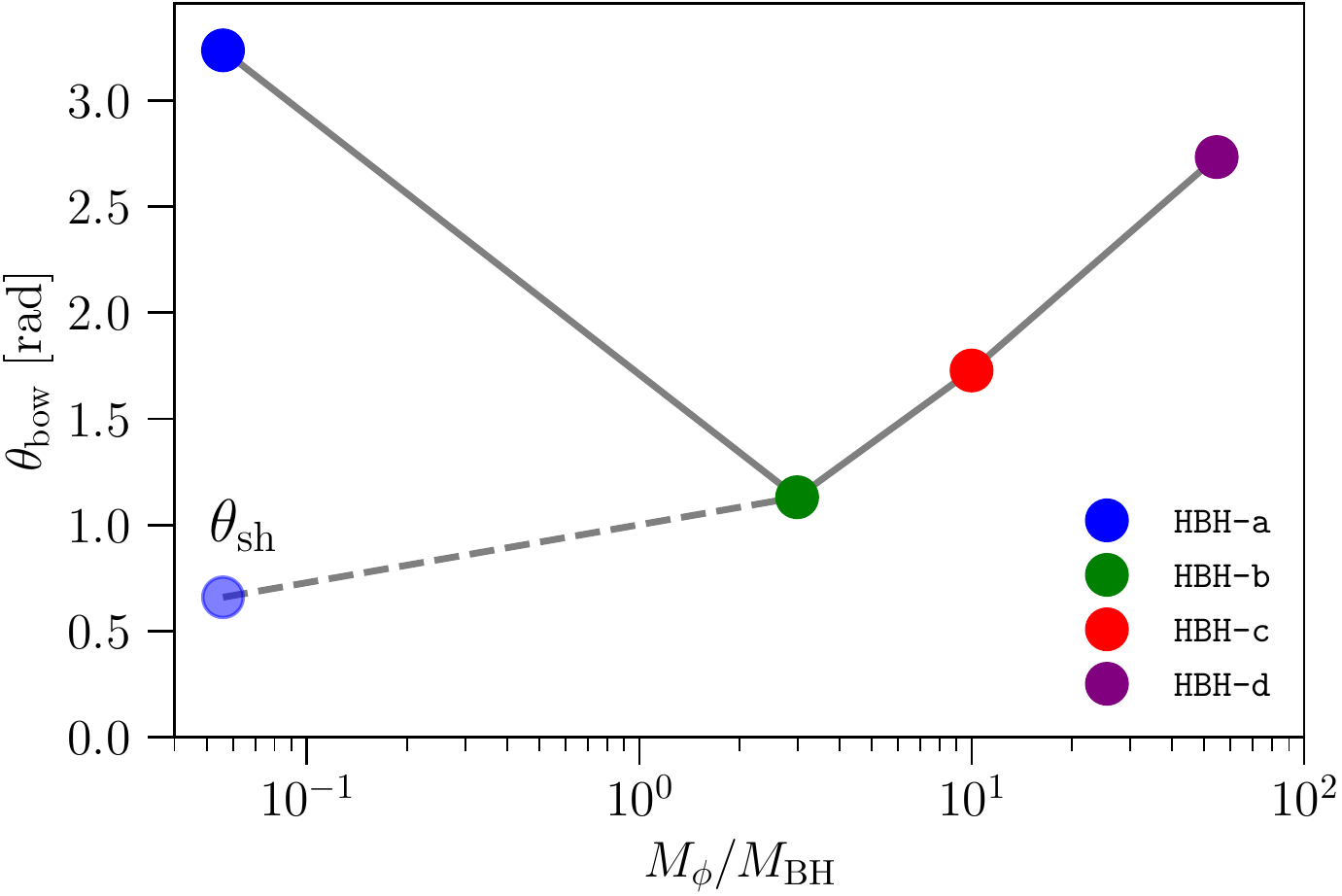}
\caption{Bow-shock opening angle $\theta_{\rm bow}$ shown as a function
  the relative weight of the scalar field via the mass ratio
  $M_{\phi}/M_{\rm BH}$ for the four HBHs considered. In the case of
  model \texttt{HBH-a} we also report the (downstream) shock-cone opening
  angle $\theta_{\rm sh}$ that shows a monotonic behaviour in terms of
  $M_{\phi}/M_{\rm BH}$.}
    \label{fig:2Drho_1}
\end{figure}

As customary in these simulations, the black hole is placed at the center
of the computational domain, which is then filled by accreting gas
modeled has having a uniform density and moving along the positive
$x$-direction. Since the coordinates used in the code are spherical, the
velocity-field components are 
\begin{eqnarray}
v^{r}   &=&  \phantom{-}v_{\infty}{\cal H}_{1} \cos \varphi + v_{\infty}{\cal H}_{2} \sin \varphi \,, \\
v^{\varphi} &=&          -v_{\infty} {\cal H}_{3} \sin \varphi  + v_{\infty}{\cal H}_{4} \cos \varphi \,, 
\end{eqnarray}
where ${\cal H}_{i}$ are functions written in terms of the metric
components and $v_{\infty}$ is the asymptotic velocity of the fluid (see
Refs.~\citep{Font1999b, Cruz2012, Cruz2013, Lora2015, Cruz2016} for more
details). On the other hand, the pressure of the gas is constructed from
the definition of the sound speed for an ideal gas, that is, $c^{2}_{\rm
  s}=p\gamma(\gamma-1)/[p\gamma-\rho(\gamma
  -1)]$~\cite{Rezzolla_book:2013}. Based on this, and using the ideal-gas
equation of state $p=\rho \epsilon(\gamma-1)$, the initial pressure of
the gas is set to
\begin{eqnarray}
p_{\rm ini} = \frac{c^{2}_{\rm s,\infty}
  \rho_{\infty}(\gamma-1)}{\gamma(\gamma-1) -c^{2}_{\rm s,\infty}\gamma}
\,,
\end{eqnarray}
where the adiabatic index $\gamma$, the speed of sound at infinity
$c_{\rm s,\infty}$, the asymptotic rest-mass density $\rho_{\infty}$ and
the asymptotic velocity $v_{\infty}$ are free parameters. Specifically,
in this work we use $\gamma=5/3$, $\rho_{\infty}=10^{-8}$, $c_{\rm
  s,\infty}=0.1$ and $v_{\infty}=0.5$. {As a result, the gas is
  supersonic, with Newtonian Mach number at infinity ${\cal
    M}:=v_{\infty}/c_{\rm s,\infty}=5$.}

All simulations reported in this work were carried out in two spatial
dimensions using the \texttt{BHAC} code~\cite{Porth2017, Olivares2018a,
  Olivares2019}, employing spherical coordinates {$(r,\varphi)$} on the
equatorial plane and {a} logarithmic radial coordinate $R:={\rm
  log}(r)$. The numerical grid over which the hydrodynamical equations 
  are solved covers the spherical-polar coordinates domain $r/r_{\rm EH}\in [1.1,
    1000]$ and $\varphi \in [0,2\pi]$; hence, the hydrodynamical grid
  always includes the grid over which the HBH solutions are computed. In
  the intermediate region, e.g., for $200 \lesssim r/M_{\rm BH} \lesssim
  1000$, we simply employ the Kerr solution in Boyer-Lindquist
  coordinates, since the differences of the HBH solutions are very small
  at these distances. In the left hemisphere (i.e., for $x<0$) we inject
  gas with constant velocity and density at the outer radius, using the
  same prescription as the initial conditions. On the other hand, in the
  right hemisphere (i.e., for $x\geq0$), we impose an outflow boundary at
  the inner and outer radii for $x>0$.  Periodic boundary conditions are
imposed in the {$\varphi$} direction. {For the simulations} we use an
adaptive mesh refinement algorithm~\cite{Loehner87}, with three levels of
refinement, where the coarse grid resolution corresponds to $\Delta
R=3.56 \times 10^{-3}$ for the logarithmic radial coordinate and $\Delta
\varphi/\pi = 3.9 \times 10^{-3}$ for the angular one. Additionally, we
use the finite-volume method {to} solve the general relativistic
hydrodynamics equations \eqref{eq:conservationlaw}, {with the particular
  choices of the HLLE (Harten, Lax, van Leer, and Einfeldt) flux
  formula~\cite{Harten83,Einfeldt88} and the \textit{minmod} slope
  limiter to compute the fluxes and reconstruct the primitive variables
  at cell interfaces, respectively~\cite{Rezzolla_book:2013}. The time
  update in all simulations is} performed using a third-order Runge-Kutta
time integrator.

\section{Results}
 \label{sec:results}
\begin{figure*}
\centering
\includegraphics[width=1.0\textwidth]{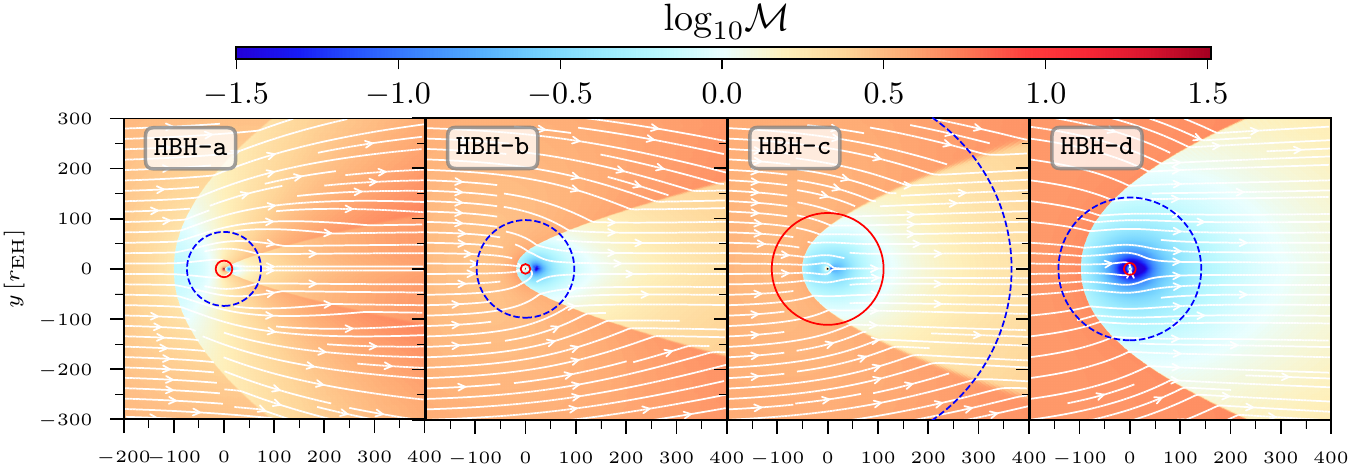}
\includegraphics[width=1.0\textwidth]{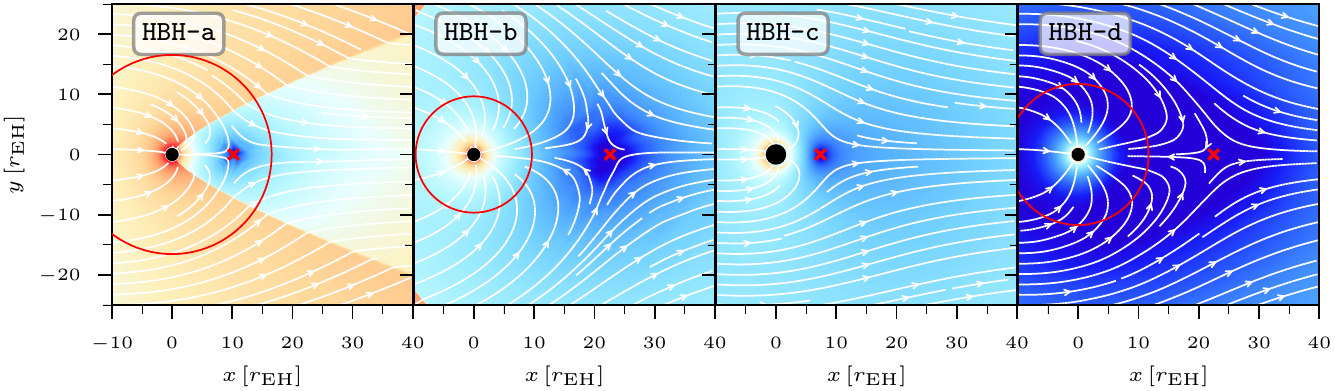}
\caption{The same as in Fig.~\ref{fig:2Drho} but for the Newtonian Mach
  number ${\cal M}$. In addition, shown with white lines are the fluid
  streamlines, while a red cross is used to mark the position of the
  stagnation point.} \label{fig:2DMach}
\end{figure*}

The morphology of the accreting gas after the simulations have reached a
steady-state is shown in the different panels of Fig.~\ref{fig:2Drho} for
all HBHs of our sample. The figure displays with a colormap and
isocontours the two-dimensional distribution of the rest-mass density at
$t=10000\,M_{\rm BH}$ and normalised to its asymptotic value.  The
morphology of the various configurations exhibits the typical behavior of
relativistic BHL accretion onto a moving black hole. More specifically,
and as first learnt in simulations involving Kerr black holes (see, e.g.,
Refs.~\cite{Font1999b, Donmez2010, Cruz2012, Lora2015219, Cruz2020b}),
when a supersonic gas moves past an HBH, a shock-cone appears downstream,
in the proximity of the black hole. This shock-cone is clearly visible
not only in Fig.~\ref{fig:2Drho}, but also in the left panels of
Fig.~\ref{fig:2DMach}, which displays instead the distribution of the
Mach number. Each panel in Fig.~\ref{fig:2Drho} reports with red solid
circles the location of the maximum scalar field $\phi_{\rm max}$, while
with blue dashed circles the outer edge of the torus of scalar hair,
which is defined as the radius where the normalized scalar field density
is five orders of magnitudes smaller than $\phi_{\rm max}$. Hence the
region between the two circles can be taken as an indication of the size
of the torus of scalar hair.

Despite these common features, each HBH solution interacts with the
inflowing plasma developing different gas patterns. Such differences can
be observed in the shock-cones temperature profiles, velocity fields (see
streamlines in Fig.~\ref{fig:2DMach}), and opening angles $\theta_{\rm
  sh}$. The latter are plotted in Fig.~\ref{fig:2Drho_1} and are measured
as follows. We compute the normalized rest-mass density as a function of
the $\varphi$ coordinate at $r=200\,r_{\rm EH}$ and then mark the opening
angle as the angle at which the density gradient in the angular
direction, $\partial _{\varphi} \rho$, diverges; the opening angles for
the different HBH models are listed in Table~\ref{tab:results}. As shown
in Fig.~\ref{fig:2Drho_1} and reported in Table~\ref{tab:results}, the
opening angle overall increases as the ``hairiness" of the black hole is
also increased. In the case of model \texttt{HBH-a} the figure also
reports the (downstream) shock-cone opening angle $\theta_{\rm sh}$ that
shows a monotonic behaviour in terms of $M_{\phi}/M_{\rm BH}$. An
intuitive explanation of this behaviour comes when considering that as
the contribution in mass and angular momentum of the scalar hair becomes
comparable to that of the black hole, the ``local'' curvature
will be less severely modified by the black hole and hence the impact of
the black hole on the incoming flow, namely the shock-cone, will be smaller.

The HBH solution \texttt{HBH-a} corresponds to the least hairy black hole
of our set and is thus more Kerr-like. In this case, two shock fronts
develop in the flow: first, a \textit{broad} and \textit{strong}
bow-shock appears upstream of the black hole with an opening angle
$\varphi_{\rm bow} \sim \pi ~{\rm rad}$; second, a \textit{narrow} and
\textit{weak} shock-cone forms downstream, similar to what is observed for
pure Kerr black holes. When considering more hairy black holes, on the
other hand, the trailing shock-cone starts enveloping entirely the black
hole as the amount of scalar hair increases (see the transition from
model \texttt{HBH-b} to model \texttt{HBH-d}), transforming itself into a
leading bow-shock. The explanation for this different flow morphology has
to be found again in the fact that the curvature, especially near the
black hole, becomes less severe as the strength of the scalar field is
increased, so that the impact of the black hole on the incoming flow is
less and less marked. As a result, the second downstream shock weakens up
to disappearing, as can be seen in the various panels of
Fig.~\ref{fig:2Drho}.

Additionally, the different HBH solutions explored in our simulations
have different accretion properties, which can be characterized in
terms of several quantities. The first one is the so-called accretion
radius, defined as~\cite{Rezzolla_book:2013}
\begin{equation}
r_{\rm acc}:= \frac{M_{\rm BH}}{c^{2}_{s, \infty} + v^{2}_{\infty}}\,,
\end{equation}
and which marks the transition of the flow from subsonic to supersonic
velocities. The accretion radius can also be interpreted as marking the
spherical region within which gravity dominates and so it is expected to
decrease as the amount of the scalar hair is increased [cf.,
  Table~\ref{tab:results}]. Other important quantities are the accretion
rates of rest-mass $\Dot{M}$ and linear momentum\footnote{Since our
incoming flow does not have any angular momentum, no angular momentum can
be transferred to the black hole, at least in the test-fluid
approximation. We have verified that this is the case by measuring the
flux of angular momentum at the event horizon.}  $\Dot{P}^i$, which are
computed as~\citep{Petrich89}
\begin{eqnarray}
\label{eq:mdot}
\Dot{M}    &:=& \int_{0}^{2\pi} \alpha\sqrt{ \tilde{\gamma} }D(v^{r} - \beta^{r}/\alpha) d\varphi\,,\\
\Dot{P}^{r} &:=&-\int_{\partial V} \alpha \sqrt{\tilde{\gamma}} T^{r j} d\Sigma_{j}
+ \int_{V} \alpha \sqrt{\tilde{\gamma}} S^{r} dV\, ,
\label{eq:pdot}
\end{eqnarray}
where the inward fluxes are measured at the event horizon of each HBH
solution. The volume and surface integrals were calculated at volume $V$
and surface $\partial V$ defined by the sphere with radius $2 r_{\rm
  EH}$. In the case of spherical accretion onto nonrotating black holes,
$\Dot{M}$ and $\Dot{P}^i$ can be computed analytically to
be~\citep{Petrich89, Rezzolla_book:2013}
\begin{eqnarray}
  \label{eq:canmdot}
  \dot{M}_{\rm ref}&=& 4 \pi \lambda \rho_{\infty} M_{\rm BH}^{1/2}r^{3/2}_{\rm
    acc}\,,\\
  \label{eq:canpdot}
  \dot{P}_{\rm ref}&=& \dot{M}_{\rm ref}\frac{v_{\infty}}{\sqrt{1-v^{2}_{\infty}}}\,,
\end{eqnarray}
where
\begin{equation}
\lambda := \left(\frac{1}{2}\right)^{(\gamma +1)/(2(\gamma -1))}
\left(\frac{5-3\gamma}{4}\right)^{-(5-3\gamma)/(2(\gamma-1))}\,,
\end{equation}
with $\lambda \sim 1/4$ when $\gamma=5/3$. We will use expressions
\eqref{eq:canmdot} and \eqref{eq:canpdot} as reference values against
which we normalize the accretion rates in our set of HBH spacetimes and
report them in Table~\ref{tab:results}.

However, before discussing such quantitative properties of the accretion
flow, it is useful to return to the more general morphological
properties. To this scope, we note that the motion of the gas can be best
followed through the streamlines plotted in white in the various panels of
Fig.~\ref{fig:2DMach}, which also reports with a colormap the Mach number
in the steady-state configuration of the flow at $t=10000\,M$. Starting
from a constant velocity flow in the positive $x$-direction with
$v_{\infty}=0.5$ (note that the streamlines are parallel), we observe
that the bulk of the flow close to the black hole is accreted, while the
flow further away is hardly perturbed by the presence of the black hole
and escapes in its motion. The gas that is captured experiences a sharp
transition from a supersonic flow to a subsonic one as it is braked by
the black hole and subsequently accreted. The Mach number reaches
particularly small values for solutions where the mass and spin of the
scalar field dominate over those of the black hole, namely for model
\texttt{HBH-d}. Interestingly, as the scalar hair of the black hole
increases, the velocity distribution becomes more and more spherically
symmetric, as can be appreciated when contrasting models \texttt{HBH-a}
and \texttt{HBH-d}. 

\begin{figure}
\centering
\includegraphics[width=0.5\textwidth]{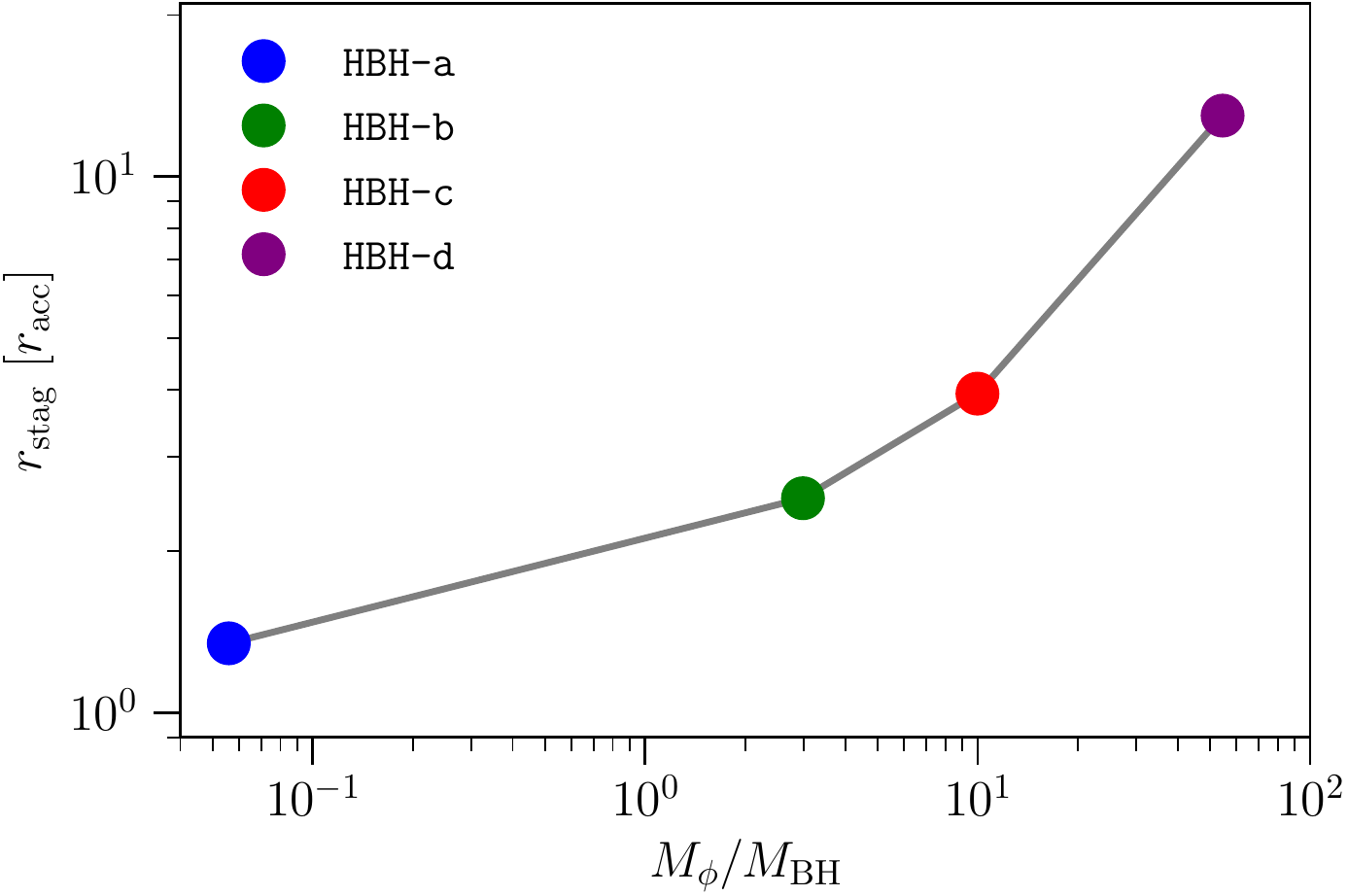}
\caption{Radial position of the stagnation point in units of the
  accretion radius and shown as a function the relative weight of the
  scalar field via the mass ratio $M_{\phi}/M_{\rm
    BH}$.} \label{fig:2DMach_1}
\end{figure}

Figure~\ref{fig:2DMach} also reveals that there are streamlines
converging into a single point downstream of the black hole and inside
the shock-cone. At such a point, which is identified in the figure with a
red cross and is commonly referred to as the ``stagnation point''
$r_{\rm stag}$, the velocity is zero and the fluid is in unstable
equilibrium, so that any perturbation can either push the fluid into the
black hole or away from it, in the downstream flow.
Figure~\ref{fig:2DMach_1} reports the location of the stagnation point as
a function of the mass ratio $M_{\phi}/M_{\rm BH}$ for our four HBH
models when expressed in terms of the accretion radius $r_{\rm acc}$.
Also in this case, it is apparent that the hairier the solution the
larger the stagnation radius and hence the more extended the low-velocity
(subsonic) region. This is also clear in the various panels of
Fig.~\ref{fig:2DMach}, where different shades in the colormap
characterize models \texttt{HBH-a} and \texttt{HBH-d}.  As for other
quantities, Table~\ref{tab:results} reports the mean value and the
standard deviation of $r_{\rm stag}$ for every HBH of our sample,
measured from the saturation time (defined below) up to the end of each
numerical simulation.

\begin{figure}
\centering
\includegraphics[width=0.49\textwidth]{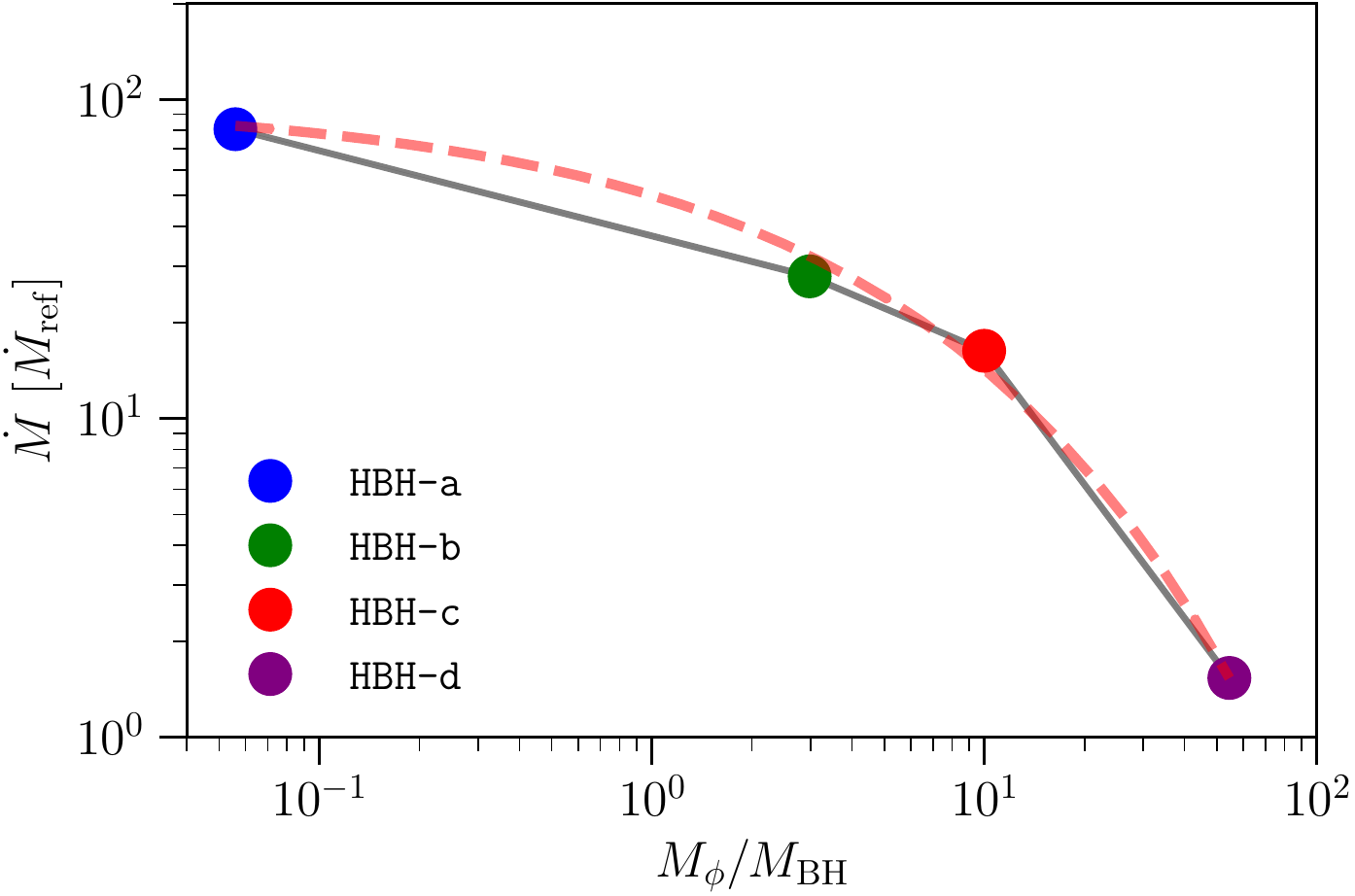}
\includegraphics[width=0.49\textwidth]{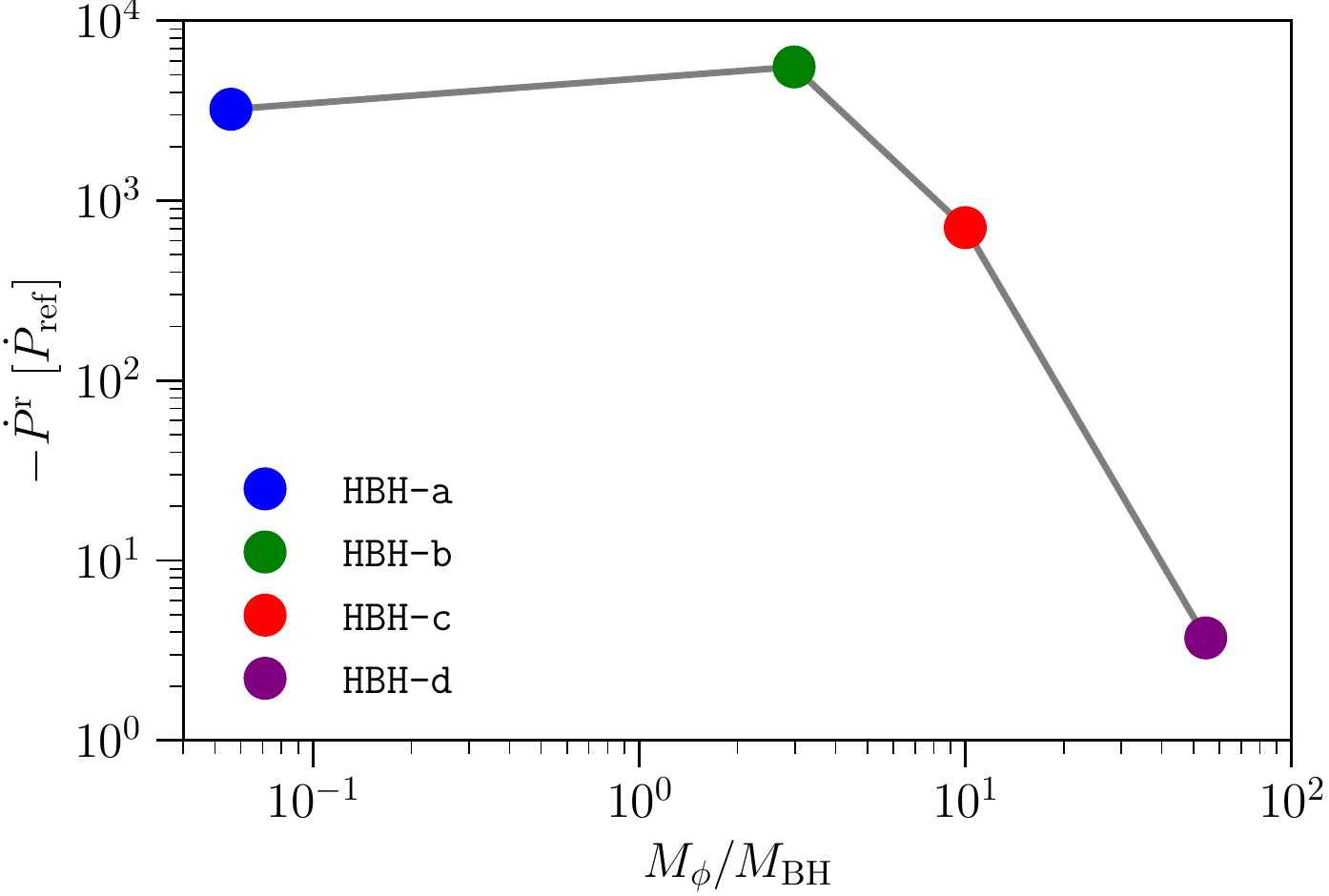}
\caption{Accretion rates of rest-mass (top panel) and of the radial
  momentum (bottom panel); both quantities are normalized by the
  corresponding quantities ($\dot{M}_{\rm ref}$, $\dot{P}_{\rm ref}$) in
  the case of spherical accretion onto a nonrotating black hole and refer
  to values attained when the flow has reached a steady-state. The red
  dashed line shows the fit given by
  expression~\eqref{eq:mdot_fit}.} \label{fig:MassRates}
\end{figure}

In Fig.~\ref{fig:MassRates} we show the rest-mass and radial-momentum
accretion rates normalized by their reference values, $\dot{M}_{\rm ref}$
and $\dot{P}_{\rm ref}$, as a function of the mass ratio $M_{\phi
}/M_{\rm BH}$. The values reported correspond to the steady-state flow
solution after the equilibrium time $\tau_{\rm eq}$ has been
achieved. The latter is defined as the time when the accretion
  process starts reaching a stationary state (i.e., when ${\ddot{M}} \le
  10^{-4}/M$) and clearly varies with the amount of scalar hair, becoming
  significantly smaller as the amount of scalar field is increased (see
Table~\ref{tab:results}). Note that rest-mass accretion rates decrease
for hairier solutions, which have a shallower gravitational potential and
thus reduce the accretion efficiency. However, HBHs have quite
generically mass-accretion rates that are larger than their corresponding
hairless counterparts.  Similarly, the radial-momentum accretion rate
also decreases from model $\texttt{HBH-a}$ to model $\texttt{HBH-d}$,
which is consistent with the fact that the amount of matter able to
transfer linear momentum decreases with the increase of scalar
field. Also for this quantity, HBHs have quite generically
momentum-accretion rates that are three or even four orders of magnitude
larger than pure Kerr black holes.

Once the system reaches a steady-state it is possible to compute
semi-analytical relations for the mass accretion rates as a function of
the scalar-field-to-black-hole mass ratio. Fits to our numerical results
in the steady-state regime are given by 
\begin{eqnarray}
\log _{10}\left(\frac{\dot{M}}{\dot{M}_{\rm ref}}\right) &=& 2 - 0.3
\left( \frac{M_{\phi}}{M_{\rm BH}}
\right)^{0.45}\,. \label{eq:mdot_fit}
\end{eqnarray}
The expression above, which is shown in the top panel of
Fig.~\ref{fig:MassRates} with a dashed line, can be used to explore, for
instance, progenitor initial masses and spins of binary black hole
gravitational-wave events~\cite{Abbott2019c,Abbott2021c,LVK2021} when
accounting for the possible effects a surrounding ultralight scalar field
might have. Efforts along these lines, albeit without the inclusion of
bosonic clouds, have been reported recently in connection with the nature
of the transient event GW190521. Using Newtonian~\cite{DeLuca2021,
  DeLuca2021b} and relativistic accretion models~\cite{Cruz2020b,
  Cruz2021a} three possible scenarios for the origin of the progenitor
have been discussed, namely involving primordial black holes,
stellar-mass black holes formed by direct collapse, and intermediate-mass
black holes resulting from the collapse of Pop III stars. Taking into
account the potential contribution of ultralight scalar field dark matter
is relevant to further assess those models.

\begin{table*}
\begin{center}
  \renewcommand{\arraystretch}{1.4}
  \setlength\tabcolsep{3.pt}
    \begin{tabular}{lcccccccc}
\hline\hline
 Model    & $r_{\rm acc}$  & $\dot{M}_{\rm ref}$ &$\dot{P}_{\rm ref}$&$\langle \dot{M} \rangle$ & $\langle \dot{P}^{r} \rangle$ & $r_{\rm stag}$ & $\theta_{\rm bow}$ & $\tau_{\rm eq}$  \\
     & $[M_{\rm BH}]$  & & & $[\dot{M}_{\rm ref}]$ & $[\dot{P}_{\rm ref}]$ & $[r_{\rm acc}]$ & $[{\rm rad}]$ & $[M_{\rm ADM}]$  \\
\hline
\hline
\texttt{HBH-a} & $1.51$ & $3.66\times 10^{-8}$  & $2.11\times 10^{-8}$  & $9.01\times 10^{1}$ & $-3.59\times 10^{3}$    &  $1.35$  & $3.24$ & $88$  \\
\texttt{HBH-b} & $0.90$ & $1.30\times 10^{-8}$  & $7.50\times 10^{-9}$  & $4.43\times 10^{2}$ & $-8.79\times 10^{4}$    &  $2.51$  & $1.13$ & $24$  \\
\texttt{HBH-c} & $0.19$ & $5.70\times 10^{-10}$ & $3.29\times 10^{-10}$ & $1.97\times 10^{3}$ & $-8.55\times 10^{4}$  &  $3.94$  & $1.73$ & $5$   \\
\texttt{HBH-d} & $0.07$ & $7.30\times 10^{-11}$ & $4.21\times 10^{-11}$ & $4.74\times 10^{3}$ & $-1.15\times 10^{4}$  &  $12.99$ & $2.73$ & $6$   \\
\hline
\end{tabular}
\caption{Summary of the quantities measured numerically results for all of
  the HBHs explored in this work. Reported are the accretion radius
  $r_{\rm acc}$, the rest-mass and radial-momentum accretion rates
  $\dot{M}$ and $\dot{P}^{r}$ expressed in terms of the corresponding
  quantities $\dot{M}_{\rm ref}$ and $\dot{P}_{\rm ref}$ for spherical
  accretion onto a nonrotating black hole, the position of the stagnation
  $r_{\rm stag}$ point in $r_{\rm acc}$ units, the opening angle of the
  bow-shock $\theta_{\rm bow}$, and the equilibrium time $\tau_{\rm eq}$
  needed for the flow to reach steady-state.}
\label{tab:results}
\end{center}
\end{table*}

\section{Application to QPOs in Sgr~A* and microquasars}
\label{sec:AA}

As an astrophysical application of the results on BHL-accretion past HBHs
reported above and to find potential associations with the QPOs observed
in Sgr A* and in microquasars, we consider the frequency spectrum of the
fluid oscillations that arise within the shock-cone and that are related
to the stagnation point. We recall that the association of QPOs with
oscillations of the shock-cone in the downstream part of a BHL flow was
first reported in Ref.~\citep{Donmez2010}. In essence, it was noted that
the shock-cone acts as a cavity trapping pressure perturbations and
exciting specific $p$-modes (see Ref.~\cite{Rezzolla_qpo_03a,
  Rezzolla_qpo_03b, Zanotti05} for a similar process taking place in
accretion disks around black holes in general relativity). For black-hole
spacetimes without scalar hair, it was found that the frequencies of
these modes depend on the black-hole spin and on the properties of the
flow, and scale linearly with the inverse of the black-hole
mass~\citep{Donmez2010}. Here, we investigate how much the presence of
the scalar field in our HBH spacetimes modifies the mode frequencies of
the shock-cone vibrations and the additional resonant cavities that
emerge because of the presence of a scalar field.

\begin{figure}
  \centering
  \includegraphics[width=0.5\textwidth]{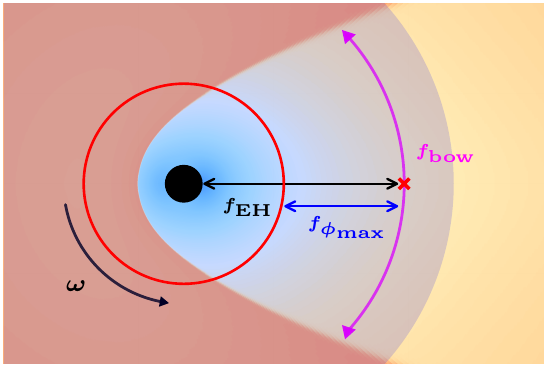}
  \caption{Schematic diagram of the resonant cavities produced in a BHL
    flow onto a HBH where a $p$-mode can be trapped.  The first cavity is
    formed between the stagnation point (red cross) and the event
    horizon, with an associated frequency $f_{\rm EH}$ (black arrows). The
    second one is between the stagnation point and the position of the
    maximum of the scalar field amplitude (red solid line) with frequency
    in this cavity is $f_{\phi_{\rm max}}$ (blue arrows), while the
    scalar field orbital frequency is $\omega$. The third cavity is that
    enclosed by the bow-shock where $p$-modes travel transversally to the
    shock-cone with frequency $f_{\rm bow}$ (magenta arrows). Finally, in the
    case of model \texttt{HBH-a} (and of a Kerr black hole) another
    cavity is present in the downstream shock-cone with $p$-modes trapped
    at frequency $f_{\rm sh}$. The color map shows the rest-mass density
    of the gas and the ultralight scalar hair distribution (grey-shaded
    area). 
    }
\label{fig:Freqs}
\end{figure}

Figure \ref{fig:Freqs} provides a schematic diagram of the possible
resonant cavities that can be formed around the stagnation point, and
where pressure perturbations can propagate a distance $L$ with velocity
$c_{s}$. To cover all of these possibilities, we define a generic
resonant $p$-mode frequency as $f_{i} := c_{s, i}/L_{i}$, where $i$ is
not an integer and is used to indicate the different resonant cavities
that can be produced in the flow (see Table~\ref{tab:QPOs}). The first
cavity is then formed between the stagnation point (red cross) and the
event horizon, with an associated frequency $f_{\rm EH}$ (black
arrows). The second one is between the stagnation point and the position
of the maximum of the scalar field amplitude (red solid line) whose
rotation frequency is $\omega$; the $p$-mode oscillations in this cavity
have a frequency $f_{\phi_{\rm max}}$ (blue arrows). The third cavity is that
enclosed by the upstream bow-shock and in this region sound waves travel
transversally to the shock with frequency $f_{\rm bow}$ (red
arrows). Finally, in the case of model \texttt{HBH-a} (and of a Kerr
black hole) another cavity is present in the downstream shock-cone with
$p$-modes trapped at frequency $f_{\rm sh}$.

The resonant frequencies expected on the basis of the measured values of
$L_i$ and $c_{s,i}$ are reported for each HBH in Table~\ref{tab:QPOs} and
in units of $1/M_{\rm ADM}$. Note that the frequencies at the shock-cone
are different for each HBH model because, on the one hand, the opening
angle increases with the mass ratio $M_{\rm SF}/M_{\rm BH}$ (see
Fig.~\ref{fig:2Drho_1}) and, on the other hand, the sound speed also
changes when the gas moves from supersonic to subsonic
velocities. Similarly, the frequency $f_{\rm EH}$ is also different for
each HBH since the position of the stagnation point changes as a function
of $M_{\phi}/M_{\rm BH}$ (see Fig.~\ref{fig:2DMach_1}).

\begin{figure}
\centering
\includegraphics[width=0.50\textwidth]{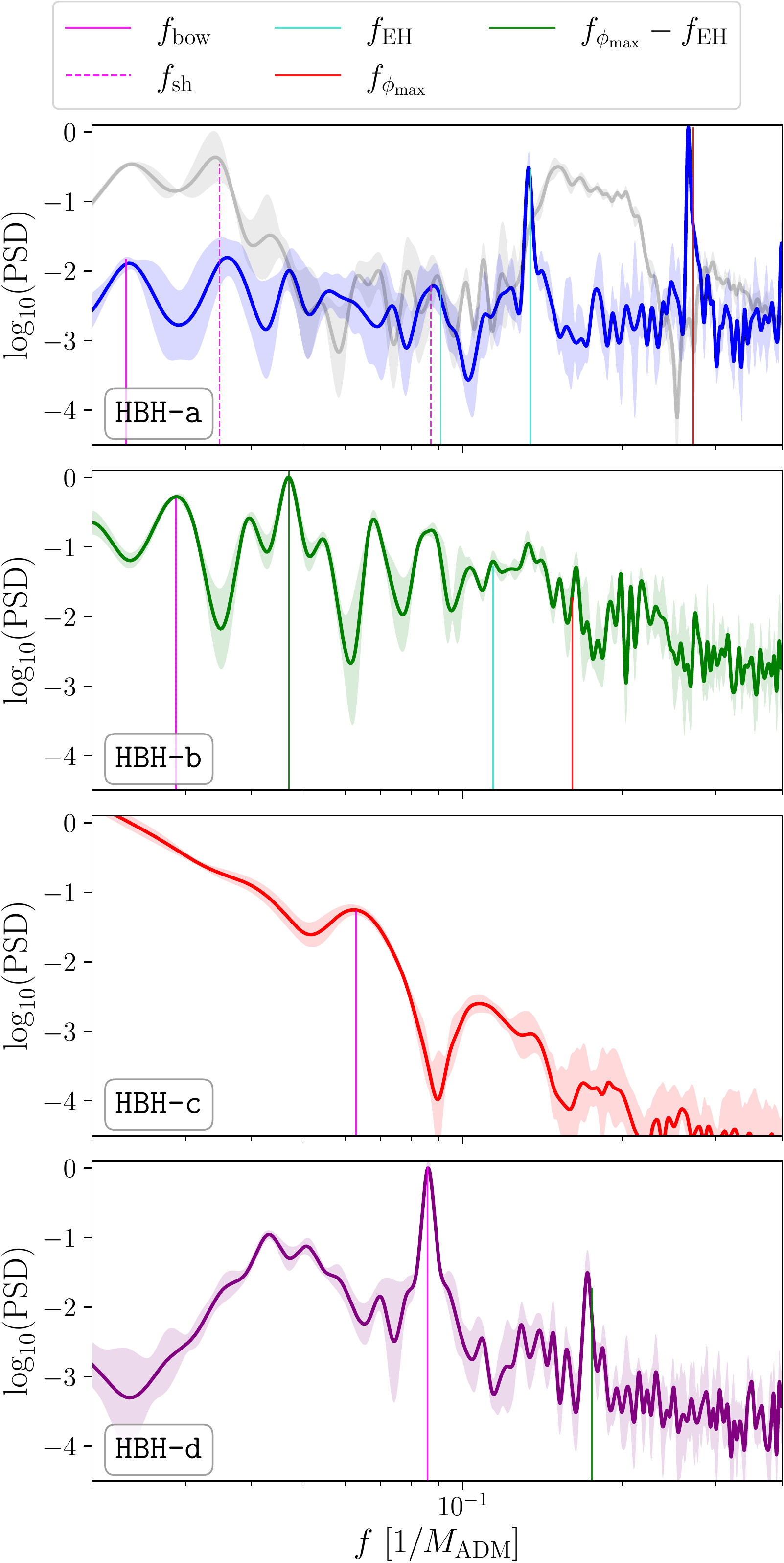}
\caption{Power spectral densities (PSD) of the rest-mass density
  evolution measured at the stagnation point. Each panel refers to a
  different HBH, with colored solid lines showing the PSDs and the
  corresponding shaded regions indicating the standard deviation. The
  grey line and area reports the PSD for a Kerr black hole with spin
  $a_{\star}=15/16$. The expected values of the representative
  frequencies are reported with colored vertical lines (see also
  Table~\ref{tab:QPOs}). }
\label{fig:Freqs2}
\end{figure}

The predictions for the expected frequencies can be compared with the
power spectral density (PSD) as obtained after performing a Fourier
transform of the rest-mass density measured at the stagnation point of
each HBH model. Furthermore, to increase the robustness of our results,
we also compute the PSDs in ten detectors located around the stagnation
point and compute the mean and the corresponding standard deviation of
the measured frequencies. Figure \ref{fig:Freqs2} shows the mean PSD
(colored solid lines) and the standard deviation for each HBH spacetime
(colored shaded regions). In addition, the grey line and shaded region in
the top panel for model \texttt{HBH-a} corresponds to a Kerr black hole
with spin $a_{\star}=15/16\ (\Omega_{\rm EH}=0.35)$, which we use as a
reference\footnote{We consider a highly rotating Kerr black hole because
our HBH solutions all have high spins. Note also that the dimensionless
spin of the HBH solutions, $a_{\rm BH}:=J_{\rm BH}/M_{\rm BH}^2$, is not
bounded by $\pm 1$ (see Table~\ref{tab:1}). Finally, the rotation of the
scalar field is synchronized with the black hole, {\it i.e}, the
nonrotating solution reduces to Schwarzschild with zero scalar field.}

\begin{figure}
  \centering
  \includegraphics[width=0.50\textwidth]{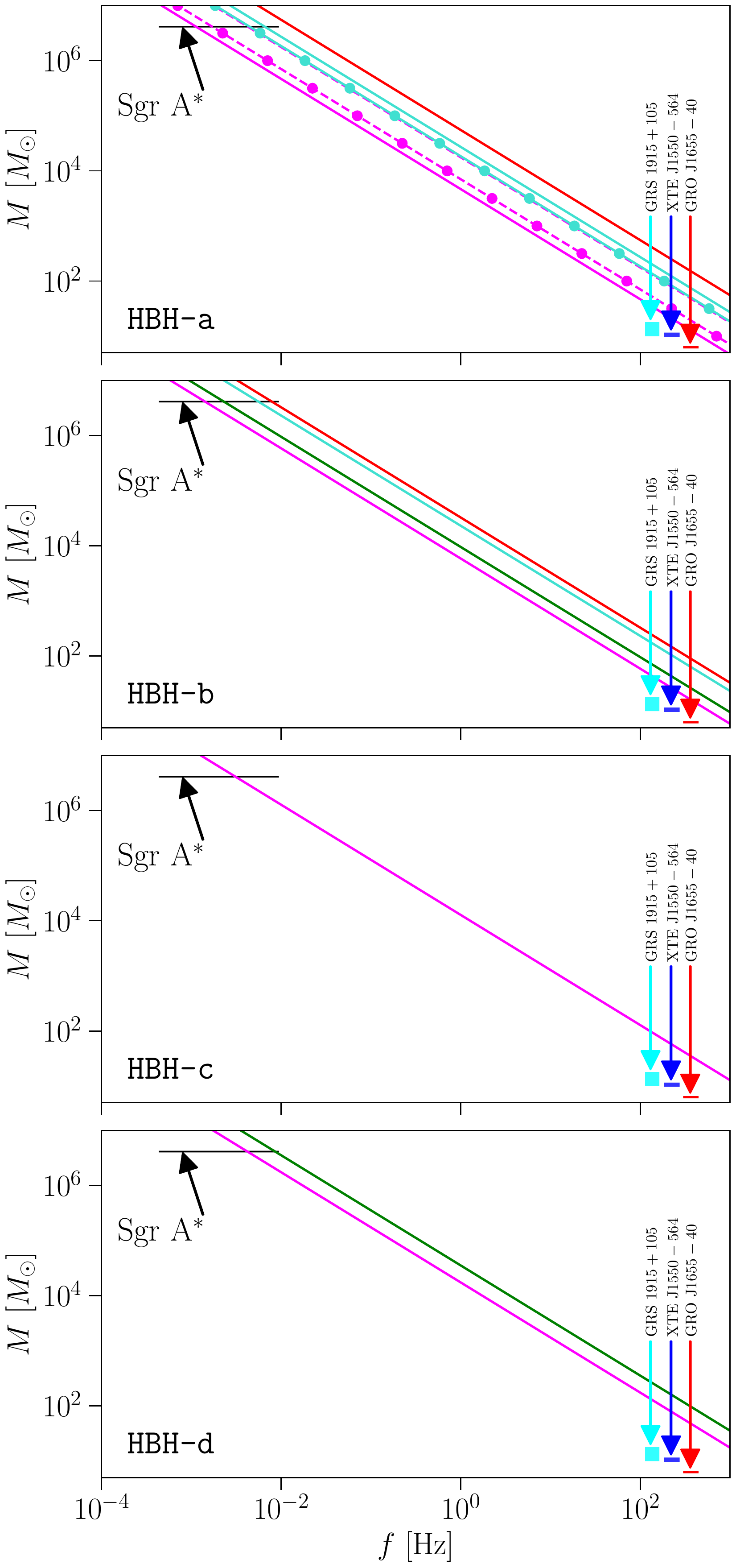}
\caption{The four panels report the QPO frequencies for the four HBHs
  considered and black-hole masses in the range $M/M_{\odot}\in
  [1,\,10^{7}]$. The colored lines correspond to the vertical lines shown
  in Fig.~\ref{fig:Freqs2}, while the filled markers in the top panel
  refer to the QPO frequencies for a Kerr black hole. Also reported are
  the observed QPO frequencies for Sgr~A* $(f/{\rm mHz} \in
  {\rm[0.434,\,10]})$, {GRO J1655-40} $(f/{\rm mHz} \in {\rm [300,\,
      450]}$), {XTE J1550-564} at 184\,Hz and 276\,Hz, and for {GRS
    1915+105} $(f/{\rm mHz} \in {\rm [113,\, 168]})$.}
\label{fig:Freqs3}
\end{figure}

For all models, we identify several peaks in the
spectrum and use vertical lines to report the predicted resonant
frequencies described above. Starting from the reference pure-Kerr black
hole, it is possible to recognize in the PSD the low-frequency $p$-mode
in the shock-cone $f_{\rm sh}$ (together with overtones at $2f_{\rm sh}$,
$3f_{\rm sh}$, and $5f_{\rm sh}$) and the high-frequency resonance
between the accretion radius and the event horizon $f_{\rm EH}$. When
passing to the HBH solutions, it is possible to recognise for model
\texttt{HBH-a} the peaks corresponding to the $p$-mode
frequencies in the cavity produced by the (upstream) bow-shock $f_{\rm
  bow}$, the mode at $f_{\rm sh}$ trapped in the (downstream) shock-cone
(present only for model \texttt{HBH-a}), the mode at $f_{\rm EH}$ in the
cavity between the event horizon and the stagnation point, and the mode
at $f_{\phi_{\rm max}}$ in the cavity between the stagnation point and
the maximum of the scalar field. For model \texttt{HBH-b} we can clearly
identify the main peaks with frequencies at $f_{\rm bow}$, $f_{\rm EH}$,
and $f_{\phi_{\rm max}}$, but also some minor peaks which are related to
combinations of the main frequencies, e.g., $(f_{\rm bow} + f_{\rm
  EH})/2$, $3f_{\rm bow}$, and $(f_{\phi_{\rm max}} + f_{\rm EH})/2$ (not
reported in the panel). On the other hand, for model \texttt{HBH-c} we
can only identify clearly the main low-frequency mode at $f_{\rm bow}$,
while subdominant peaks appear at the first two overtones, $2f_{\rm bow}$
and $3f_{\rm bow}$. Finally, for model \texttt{HBH-d} the two dominant
peaks coincide with the trapped $p$-mode at $f_{\rm bow}$ and with the
combination of frequencies $f_{\phi_{\rm max}}-f_{\rm EH}$; this model
has the largest shock-cone opening angle and the observed low-frequency
peak corresponds interestingly to $f_{\rm bow}/2$.

\begin{table}
\begin{center}
  \renewcommand{\arraystretch}{1.4}
  \setlength\tabcolsep{3.pt}
  \begin{tabular}{l|cccccc|}
\hline\hline
${\rm Model}$  &  $f_{\rm bow}$ &$f_{\rm sh}$&$f_{\rm EH}$& $f_{\phi_{\rm max}}$ \\
\hline 
\texttt{HBH-a} & $0.023$ & $0.087$ & $0.134$ & $0.272$ \\
\texttt{HBH-b} & $0.029$ & $-$     & $0.114$ & $0.161$ \\
\texttt{HBH-c} & $0.063$ & $-$     & $0.554$ & $0.020$ \\
\texttt{HBH-d} & $0.086$ & $-$     & $0.343$ & $0.518$ \\
\texttt{Kerr}  & $-$     & $0.035$ & $0.091$ & $-$     \\
\hline
\hline
\end{tabular}
\caption{Resonance frequencies produced in the various cavities
  illustrated in Fig.~\ref{fig:Freqs}. The last row reports the frequencies for a Kerr
  black hole with $a_{\star}=15/16$; all frequencies are in units of
  $1/M_{\rm ADM}$.}
\label{tab:QPOs}
\end{center}
\end{table}

Overall, the discussion above shows that the presence of and ultralight
scalar-field does leave an imprint in the QPO frequencies of the
accreting HBH. Since each model has different frequencies, those could be
used to potentially distinguish between a black hole with a light amount
of hair (Kerr-like) from a hairier black hole (boson-star-like). Going
further, and notwithstanding the speculative nature of these
considerations, we could compare the resonant frequencies of our HBH
models with observational measurements of QPO frequencies in actual
astronomical sources to see the role scalar fields might play to fit the
data. To do so, we use the QPOs observed in the X-ray lightcurve of
Sgr~A*~\cite{Torok2005} and of three microquasars\footnote{Since
  the rest-mass density and pressure are directly related to the gas
  temperature (\i.e., $T \propto p/\rho$), and the temperature regulates
  the emissivity and absorptivity of the gas, it is perfectly
  reasonable to relate the time variability of the physical properties
  with the variability in the X-ray luminosity. Indeed, this is what is
  done when modelling kilohertz QPOs in microquasars (see, e.g.,
  \cite{Schnittman06} }, namely GRO~J1655-40, XTE~J1550-564, and
GRS~1915+105~\cite{Orosz2002}. Figure \ref{fig:Freqs3} shows the main QPO
frequencies we identify in our models, using the same color coding as in
Fig.~\ref{fig:Freqs2}, as a function of the black hole mass.

Unfortunately, the outcome of this comparison is not conclusive for the
case of Sgr~A* since the range of QPO frequencies is so large that all of
our HBH models can explain the observed frequencies as due to either to
$p$-mode oscillations trapped in the bow-shock or to radial oscillations
between the stagnation point and the event horizon. We should also note
that the QPOs of a Kerr black hole are also compatible with the observed
frequencies, e.g.,~for a $M\sim 10\,M_{\odot}$ Kerr black hole the
frequency is $\sim 100\,$Hz, (see also the results
from~\citep{Donmez2010}). On the other hand, for the three microquasar
sources considered, no HBH or Kerr black hole can explain the observed
QPO frequencies. The only exception is model \texttt{HBH-a}, for which
the low-frequency QPO $f_{\rm bow}$ is low enough to be close to the
frequency observed for GRO J1655-40.

\section{Summary}
\label{sec:discussion}
 
We have studied relativistic BHL
accretion onto an illustrative sample of the hairy black-hole
solutions reported by~\cite{Herdeiro2014, Herdeiro2015}. The hair has been assumed to
be composed of an ultralight, complex scalar field, minimally coupled to
Einstein's gravity. We have considered four solutions, describing HBHs
with different amounts of hair, ranging from Kerr-like to boson
star-like. Through numerical simulations using the \texttt{BHAC} code, we
have explored the dynamics and flow morphology of a $\gamma=5/3$ ideal
gas moving supersonically past the four HBHs and searched for signatures
in the observable quantities that may provide evidence of a deviation
from the dynamics expected from Kerr back holes.

The simulations have revealed that the morphology of the flow past HBHs
is fairly similar to that found when studying relativistic BHL accretion
onto a Kerr black hole~\citep{Font1999b, Donmez2010, Cruz2012,
  Lora2015219, Cruz2020b}. More specifically, all simulations lead to a
steady-state morphology characterized by the formation of a bow-shock
cone in the upstream part of the flow, where the gas becomes subsonic,
along with the presence of a stagnation point where the flow velocity
vanishes. However, as the amount of hair around the black hole increases,
so does the opening angle of the bow-shock-cone, while the stagnation
point moves further downstream and away from the black hole. Both of
these effects can be clearly associated to the presence of the scalar
field that distorts the spacetime near the black hole and reduces
locally the curvature. The weakening of the gravitational influence of the
black hole as a result of the increase of the scalar-hair strength has an
additional manifestation in the so-called shock-cone that develops for
BHL onto black holes in general relativity and hence without scalar
hair~\citep{Font1999b, Donmez2010}. This very specific signature that is
very robust in hairless black holes is recovered in our analysis only in
the less hairy of the HBHs (\texttt{HBH-a}) and is absent (or is very
weak) as the amount of scalar hair is increased and the overall
compactness of the spacetime decreased (the gravitational well is
distributed over a larger volume).

Our simulations have also allowed to compute the rest-mass and
linear-momentum accretion rates, finding that they attain stationary
values after a (model-dependent) equilibration time is reached. A
semi-analytical fit for the rest-mass accretion rate in terms of the mass
ratio $M_{\phi}/M_{\rm BH}$ has been found, which could potentially be
used to analyze different black-hole formation scenarios in the presence
of ultralight scalar-field dark matter.

Finally, we have explored how the modifications in the spacetime affect
one of the most intriguing aspects of BHL accretion onto black holes,
namely the presence of quasi-periodic oscillations corresponding to
resonant waves trapped in the cavities that are produced in the flow. The
existence of such QPOs, which was first pointed out in
Ref.~\citep{Donmez2010} in the case of a Kerr black hole, has been
confirmed also in the case of the HBHs considered here, where they appear
with a much richer phenomenology. This is because BHL accretion onto HBHs
leads to the formation of a number of resonant cavities, namely: in the
upstream bow-shock and, for weak scalar fields, in the downstream
shock-cone, but also between the stagnation point and the event horizon,
and between the stagnation point and the position of the maximum of the
scalar field amplitude.  We have computed such frequencies and have
compared them with the QPOs observed in the X-ray light curve of Sgr~A*
and three microquasars. Although the results of this comparison are
  not conclusive, possibly because the galactic centre is a highly
  complex and dynamical system with extremely low accretion rates and
  with inflow geometries that are not necessarily those of classical disc
  accretion, we argue that further work along these lines and the
  measurement of QPOs in other supermassive black holes might help
  establish a possible observational fingerprint for ultralight scalar
  field dark matter.

In summary, the results presented here provide a first quantitative
description of the BHL gas dynamics past HBHs, where the gravitational
field is produced by the combined presence of a black hole and a scalar
hair. In an astrophysical context, this study can help associate specific
features in the evolution of the gas to different system components,
i.e., either the black hole, the scalar field, or the combination of the
two through their gravitational interaction.

\acknowledgments

ACO and LR gratefully acknowledge the European Research Council for the
Advanced Grant ``JETSET: Launching, propagation and emission of
relativistic jets from binary mergers and across mass scales'' (Grant
No. 884631).
JAF is supported by the Spanish Agencia Estatal de Investigaci\'on
(PGC2018-095984-B-I00 and PID2021-125485NB-C21 funded by
MCIN/AEI/10.13039/501100011033 and ERDF A way of making Europe) and by
the Generalitat Valenciana (PROMETEO/2019/071).
FDL-C was supported by the Vicerrector\'ia de Investigaci\'on y Extensi\'on -
Universidad Industrial de Santander, under Grant No. 3703.
LR acknowledges the Walter Greiner Gesellschaft zur F\"orderung der
physikalischen Grundlagenforschung e.V. through the Carl W. Fueck
Laureatus Chair.
This work is supported by the Center for Research and Development in
Mathematics and Applications (CIDMA) and by the Centre of Mathematics
(CMAT) through the Portuguese Foundation for Science and Technology (FCT
- Fundac\~ao para a Ci\^encia e a Tecnologia), references
UIDB/04106/2020, UIDP/04106/2020, UIDB/00013/2020 and UIDP/00013/2020.
We acknowledge support from the projects CERN/FIS-PAR/0027/2019,
PTDC/FIS-AST/3041/2020, CERN/FIS-PAR/0024/2021 and 2022.04560.PTDC. This
work has further been supported by the European Union's Horizon 2020
research and innovation (RISE) programme H2020-MSCA-RISE-2017 Grant
No. FunFiCO-777740 and by the European Horizon Europe staff exchange (SE)
programme HORIZON-MSCA-2021-SE-01 Grant No. NewFunFiCO-101086251.
The simulations were performed on the Lluis Vives cluster at the
Universitat de Val\`encia and on the Iboga cluster at Goethe University,
Frankfurt.


\bibliographystyle{apsrev} 

\begin{thebibliography}{101}
\expandafter\ifx\csname natexlab\endcsname\relax\def\natexlab#1{#1}\fi
\expandafter\ifx\csname bibnamefont\endcsname\relax
  \def\bibnamefont#1{#1}\fi
\expandafter\ifx\csname bibfnamefont\endcsname\relax
  \def\bibfnamefont#1{#1}\fi
\expandafter\ifx\csname citenamefont\endcsname\relax
  \def\citenamefont#1{#1}\fi
\expandafter\ifx\csname url\endcsname\relax
  \def\url#1{\texttt{#1}}\fi
\expandafter\ifx\csname urlprefix\endcsname\relax\def\urlprefix{URL }\fi
\providecommand{\bibinfo}[2]{#2}
\providecommand{\eprint}[2][]{\url{#2}}

\bibitem[{\citenamefont{{Event Horizon Telescope Collaboration}
  et~al.}(2019)\citenamefont{{Event Horizon Telescope Collaboration},
  {Akiyama}, {Alberdi}, {Alef}, {Asada}, {Azulay}, {Baczko}, {Ball},
  {Balokovi{\'c}}, {Barrett} et~al.}}]{EHT_M87_PaperI}
\bibinfo{author}{\bibnamefont{{Event Horizon Telescope Collaboration}}},
  \bibinfo{author}{\bibfnamefont{K.}~\bibnamefont{{Akiyama}}},
  \bibinfo{author}{\bibfnamefont{A.}~\bibnamefont{{Alberdi}}},
  \bibinfo{author}{\bibfnamefont{W.}~\bibnamefont{{Alef}}},
  \bibinfo{author}{\bibfnamefont{K.}~\bibnamefont{{Asada}}},
  \bibinfo{author}{\bibfnamefont{R.}~\bibnamefont{{Azulay}}},
  \bibinfo{author}{\bibfnamefont{A.-K.} \bibnamefont{{Baczko}}},
  \bibinfo{author}{\bibfnamefont{D.}~\bibnamefont{{Ball}}},
  \bibinfo{author}{\bibfnamefont{M.}~\bibnamefont{{Balokovi{\'c}}}},
  \bibinfo{author}{\bibfnamefont{J.}~\bibnamefont{{Barrett}}},
  \bibnamefont{et~al.}, \bibinfo{journal}{Astrophys. J. Lett.}
  \textbf{\bibinfo{volume}{875}}, \bibinfo{eid}{L1} (\bibinfo{year}{2019}).

\bibitem[{\citenamefont{{Event Horizon Telescope Collaboration}
  et~al.}(2022)\citenamefont{{Event Horizon Telescope Collaboration},
  {Akiyama}, {Alberdi}, {Alef}, {Algaba}, {Anantua}, {Asada}, {Azulay}, {Bach},
  {Baczko} et~al.}}]{EHT_SgrA_PaperI}
\bibinfo{author}{\bibnamefont{{Event Horizon Telescope Collaboration}}},
  \bibinfo{author}{\bibfnamefont{K.}~\bibnamefont{{Akiyama}}},
  \bibinfo{author}{\bibfnamefont{A.}~\bibnamefont{{Alberdi}}},
  \bibinfo{author}{\bibfnamefont{W.}~\bibnamefont{{Alef}}},
  \bibinfo{author}{\bibfnamefont{J.~C.} \bibnamefont{{Algaba}}},
  \bibinfo{author}{\bibfnamefont{R.}~\bibnamefont{{Anantua}}},
  \bibinfo{author}{\bibfnamefont{K.}~\bibnamefont{{Asada}}},
  \bibinfo{author}{\bibfnamefont{R.}~\bibnamefont{{Azulay}}},
  \bibinfo{author}{\bibfnamefont{U.}~\bibnamefont{{Bach}}},
  \bibinfo{author}{\bibfnamefont{A.-K.} \bibnamefont{{Baczko}}},
  \bibnamefont{et~al.}, \bibinfo{journal}{Astrophys. J. Lett.}
  \textbf{\bibinfo{volume}{930}}, \bibinfo{eid}{L12} (\bibinfo{year}{2022}).

\bibitem[{\citenamefont{{Remillard} and {McClintock}}(2006)}]{Remillard2006}
\bibinfo{author}{\bibfnamefont{R.~A.} \bibnamefont{{Remillard}}}
  \bibnamefont{and} \bibinfo{author}{\bibfnamefont{J.~E.}
  \bibnamefont{{McClintock}}}, \bibinfo{journal}{Ann. Rev. Astron. Astroph.}
  \textbf{\bibinfo{volume}{44}}, \bibinfo{pages}{49} (\bibinfo{year}{2006}),
  \eprint{arXiv:astro-ph/0606352}.

\bibitem[{\citenamefont{{Abbott} et~al.}(2016)\citenamefont{{Abbott}, {Abbott},
  {Abbott}, {Abernathy}, {Acernese}, {Ackley}, {Adams}, {Adams}, {Addesso},
  {Adhikari} et~al.}}]{Abbot2016-GW-detection-prl}
\bibinfo{author}{\bibfnamefont{B.~P.} \bibnamefont{{Abbott}}},
  \bibinfo{author}{\bibfnamefont{R.}~\bibnamefont{{Abbott}}},
  \bibinfo{author}{\bibfnamefont{T.~D.} \bibnamefont{{Abbott}}},
  \bibinfo{author}{\bibfnamefont{M.~R.} \bibnamefont{{Abernathy}}},
  \bibinfo{author}{\bibfnamefont{F.}~\bibnamefont{{Acernese}}},
  \bibinfo{author}{\bibfnamefont{K.}~\bibnamefont{{Ackley}}},
  \bibinfo{author}{\bibfnamefont{C.}~\bibnamefont{{Adams}}},
  \bibinfo{author}{\bibfnamefont{T.}~\bibnamefont{{Adams}}},
  \bibinfo{author}{\bibfnamefont{P.}~\bibnamefont{{Addesso}}},
  \bibinfo{author}{\bibfnamefont{R.~X.} \bibnamefont{{Adhikari}}},
  \bibnamefont{et~al.}, \bibinfo{journal}{Phys. Rev. Lett.}
  \textbf{\bibinfo{volume}{116}}, \bibinfo{eid}{061102} (\bibinfo{year}{2016}),
  \eprint{1602.03837}.

\bibitem[{\citenamefont{{Ferrarese} and {Ford}}(2005)}]{Ferrarese2005}
\bibinfo{author}{\bibfnamefont{L.}~\bibnamefont{{Ferrarese}}} \bibnamefont{and}
  \bibinfo{author}{\bibfnamefont{H.}~\bibnamefont{{Ford}}},
  \bibinfo{journal}{Space Science Reviews} \textbf{\bibinfo{volume}{116}},
  \bibinfo{pages}{523} (\bibinfo{year}{2005}), \eprint{astro-ph/0411247}.

\bibitem[{\citenamefont{{Kormendy} and {Ho}}(2013)}]{Kormendy2013}
\bibinfo{author}{\bibfnamefont{J.}~\bibnamefont{{Kormendy}}} \bibnamefont{and}
  \bibinfo{author}{\bibfnamefont{L.~C.} \bibnamefont{{Ho}}},
  \bibinfo{journal}{Annual Review of Astronomy and Astrophysics}
  \textbf{\bibinfo{volume}{51}}, \bibinfo{pages}{511} (\bibinfo{year}{2013}),
  \eprint{1304.7762}.

\bibitem[{\citenamefont{{The LIGO Scientific Collaboration} and {the Virgo
  Collaboration}}(2020)}]{Abbott2020c}
\bibinfo{author}{\bibnamefont{{The LIGO Scientific Collaboration}}}
  \bibnamefont{and} \bibinfo{author}{\bibnamefont{{the Virgo Collaboration}}},
  \bibinfo{journal}{Phys. Rev. Lett.} \textbf{\bibinfo{volume}{125}},
  \bibinfo{eid}{101102} (\bibinfo{year}{2020}), \eprint{2009.01075}.

\bibitem[{\citenamefont{{Bustillo} et~al.}(2021)\citenamefont{{Bustillo},
  {Sanchis-Gual}, {Torres-Forn{\'e}}, {Font}, {Vajpeyi}, {Smith}, {Herdeiro},
  {Radu}, and {Leong}}}]{Bustillo2021}
\bibinfo{author}{\bibfnamefont{J.~C.} \bibnamefont{{Bustillo}}},
  \bibinfo{author}{\bibfnamefont{N.}~\bibnamefont{{Sanchis-Gual}}},
  \bibinfo{author}{\bibfnamefont{A.}~\bibnamefont{{Torres-Forn{\'e}}}},
  \bibinfo{author}{\bibfnamefont{J.~A.} \bibnamefont{{Font}}},
  \bibinfo{author}{\bibfnamefont{A.}~\bibnamefont{{Vajpeyi}}},
  \bibinfo{author}{\bibfnamefont{R.}~\bibnamefont{{Smith}}},
  \bibinfo{author}{\bibfnamefont{C.}~\bibnamefont{{Herdeiro}}},
  \bibinfo{author}{\bibfnamefont{E.}~\bibnamefont{{Radu}}}, \bibnamefont{and}
  \bibinfo{author}{\bibfnamefont{S.~H.~W.} \bibnamefont{{Leong}}},
  \bibinfo{journal}{Phys. Rev. Lett.} \textbf{\bibinfo{volume}{126}},
  \bibinfo{eid}{081101} (\bibinfo{year}{2021}), \eprint{2009.05376}.

\bibitem[{\citenamefont{{Cruz-Osorio} et~al.}(2021)\citenamefont{{Cruz-Osorio},
  {Lora-Clavijo}, and {Herdeiro}}}]{Cruz2021a}
\bibinfo{author}{\bibfnamefont{A.}~\bibnamefont{{Cruz-Osorio}}},
  \bibinfo{author}{\bibfnamefont{F.~D.} \bibnamefont{{Lora-Clavijo}}},
  \bibnamefont{and}
  \bibinfo{author}{\bibfnamefont{C.}~\bibnamefont{{Herdeiro}}},
  \bibinfo{journal}{Journal of Cosmology and Astroparticle Physics}
  \textbf{\bibinfo{volume}{2021}}, \bibinfo{eid}{arXiv:2101.01705}
  (\bibinfo{year}{2021}), \eprint{2101.01705},
  \urlprefix\url{https://doi.org/10.1088/1475-7516/2021/07/032}.

\bibitem[{\citenamefont{{De Luca} et~al.}(2021{\natexlab{a}})\citenamefont{{De
  Luca}, {Desjacques}, {Franciolini}, {Pani}, and {Riotto}}}]{DeLuca2021}
\bibinfo{author}{\bibfnamefont{V.}~\bibnamefont{{De Luca}}},
  \bibinfo{author}{\bibfnamefont{V.}~\bibnamefont{{Desjacques}}},
  \bibinfo{author}{\bibfnamefont{G.}~\bibnamefont{{Franciolini}}},
  \bibinfo{author}{\bibfnamefont{P.}~\bibnamefont{{Pani}}}, \bibnamefont{and}
  \bibinfo{author}{\bibfnamefont{A.}~\bibnamefont{{Riotto}}},
  \bibinfo{journal}{Phys. Rev. Lett.} \textbf{\bibinfo{volume}{126}},
  \bibinfo{eid}{051101} (\bibinfo{year}{2021}{\natexlab{a}}),
  \eprint{2009.01728}.

\bibitem[{\citenamefont{{Shibata} et~al.}(2021)\citenamefont{{Shibata},
  {Kiuchi}, {Fujibayashi}, and {Sekiguchi}}}]{Shibata2021b}
\bibinfo{author}{\bibfnamefont{M.}~\bibnamefont{{Shibata}}},
  \bibinfo{author}{\bibfnamefont{K.}~\bibnamefont{{Kiuchi}}},
  \bibinfo{author}{\bibfnamefont{S.}~\bibnamefont{{Fujibayashi}}},
  \bibnamefont{and}
  \bibinfo{author}{\bibfnamefont{Y.}~\bibnamefont{{Sekiguchi}}},
  \bibinfo{journal}{Phys. Rev. D} \textbf{\bibinfo{volume}{103}},
  \bibinfo{eid}{063037} (\bibinfo{year}{2021}), \eprint{2101.05440}.

\bibitem[{\citenamefont{{Gamba} et~al.}(2022)\citenamefont{{Gamba}, {Breschi},
  {Carullo}, {Albanesi}, {Rettegno}, {Bernuzzi}, and {Nagar}}}]{Gamba2022}
\bibinfo{author}{\bibfnamefont{R.}~\bibnamefont{{Gamba}}},
  \bibinfo{author}{\bibfnamefont{M.}~\bibnamefont{{Breschi}}},
  \bibinfo{author}{\bibfnamefont{G.}~\bibnamefont{{Carullo}}},
  \bibinfo{author}{\bibfnamefont{S.}~\bibnamefont{{Albanesi}}},
  \bibinfo{author}{\bibfnamefont{P.}~\bibnamefont{{Rettegno}}},
  \bibinfo{author}{\bibfnamefont{S.}~\bibnamefont{{Bernuzzi}}},
  \bibnamefont{and} \bibinfo{author}{\bibfnamefont{A.}~\bibnamefont{{Nagar}}},
  \bibinfo{journal}{Nature Astronomy}  (\bibinfo{year}{2022}),
  \eprint{2106.05575}.

\bibitem[{\citenamefont{{Lightman} and {Shapiro}}(1977)}]{Lightman1977}
\bibinfo{author}{\bibfnamefont{A.~P.} \bibnamefont{{Lightman}}}
  \bibnamefont{and} \bibinfo{author}{\bibfnamefont{S.~L.}
  \bibnamefont{{Shapiro}}}, \bibinfo{journal}{Astrophys. J.}
  \textbf{\bibinfo{volume}{211}}, \bibinfo{pages}{244} (\bibinfo{year}{1977}).

\bibitem[{\citenamefont{{Begelman} et~al.}(2006)\citenamefont{{Begelman},
  {Volonteri}, and {Rees}}}]{Begelman2006}
\bibinfo{author}{\bibfnamefont{M.~C.} \bibnamefont{{Begelman}}},
  \bibinfo{author}{\bibfnamefont{M.}~\bibnamefont{{Volonteri}}},
  \bibnamefont{and} \bibinfo{author}{\bibfnamefont{M.~J.}
  \bibnamefont{{Rees}}}, \bibinfo{journal}{Mon. Not. R. Astron. Soc.}
  \textbf{\bibinfo{volume}{370}}, \bibinfo{pages}{289} (\bibinfo{year}{2006}),
  \eprint{astro-ph/0602363}.

\bibitem[{\citenamefont{{Volonteri}}(2010)}]{Volonteri2010}
\bibinfo{author}{\bibfnamefont{M.}~\bibnamefont{{Volonteri}}},
  \bibinfo{journal}{Astronomy and Astrophysics Reviews}
  \textbf{\bibinfo{volume}{18}}, \bibinfo{pages}{279} (\bibinfo{year}{2010}),
  \eprint{1003.4404}.

\bibitem[{\citenamefont{{Hoyle} and {Lyttleton}}(1939)}]{Hoyle1939}
\bibinfo{author}{\bibfnamefont{F.}~\bibnamefont{{Hoyle}}} \bibnamefont{and}
  \bibinfo{author}{\bibfnamefont{R.~A.} \bibnamefont{{Lyttleton}}}, in
  \emph{\bibinfo{booktitle}{Proceedings of the Cambridge Philosophical
  Society}} (\bibinfo{year}{1939}), vol.~\bibinfo{volume}{35}, p.
  \bibinfo{pages}{405}.

\bibitem[{\citenamefont{{Bondi} and {Hoyle}}(1944)}]{Bondi1944}
\bibinfo{author}{\bibfnamefont{H.}~\bibnamefont{{Bondi}}} \bibnamefont{and}
  \bibinfo{author}{\bibfnamefont{F.}~\bibnamefont{{Hoyle}}},
  \bibinfo{journal}{Mon. Not. R. Astron. Soc.} \textbf{\bibinfo{volume}{104}},
  \bibinfo{pages}{273} (\bibinfo{year}{1944}).

\bibitem[{\citenamefont{Bondi}(1952)}]{Bondi52}
\bibinfo{author}{\bibfnamefont{H.}~\bibnamefont{Bondi}}, \bibinfo{journal}{Mon.
  Not. R. Astron. Soc.} \textbf{\bibinfo{volume}{112}}, \bibinfo{pages}{195}
  (\bibinfo{year}{1952}).

\bibitem[{\citenamefont{{Hunt}}(1971)}]{Hunt1971}
\bibinfo{author}{\bibfnamefont{R.}~\bibnamefont{{Hunt}}},
  \bibinfo{journal}{Mon. Not. R. Astron. Soc.} \textbf{\bibinfo{volume}{154}},
  \bibinfo{pages}{141} (\bibinfo{year}{1971}).

\bibitem[{\citenamefont{{Matsuda} et~al.}(1987)\citenamefont{{Matsuda},
  {Inoue}, and {Sawada}}}]{Matsuda1987}
\bibinfo{author}{\bibfnamefont{T.}~\bibnamefont{{Matsuda}}},
  \bibinfo{author}{\bibfnamefont{M.}~\bibnamefont{{Inoue}}}, \bibnamefont{and}
  \bibinfo{author}{\bibfnamefont{K.}~\bibnamefont{{Sawada}}},
  \bibinfo{journal}{Mon. Not. Roy. Astr. Soc.} \textbf{\bibinfo{volume}{226}},
  \bibinfo{pages}{785} (\bibinfo{year}{1987}).

\bibitem[{\citenamefont{{Fryxell} and {Taam}}(1988)}]{Fryxell1988}
\bibinfo{author}{\bibfnamefont{B.~A.} \bibnamefont{{Fryxell}}}
  \bibnamefont{and} \bibinfo{author}{\bibfnamefont{R.~E.}
  \bibnamefont{{Taam}}}, \bibinfo{journal}{Astrophys. J.}
  \textbf{\bibinfo{volume}{335}}, \bibinfo{pages}{862} (\bibinfo{year}{1988}).

\bibitem[{\citenamefont{{Sawada} et~al.}(1989)\citenamefont{{Sawada},
  {Matsuda}, {Anzer}, {Boerner}, and {Livio}}}]{Sawada1989}
\bibinfo{author}{\bibfnamefont{K.}~\bibnamefont{{Sawada}}},
  \bibinfo{author}{\bibfnamefont{T.}~\bibnamefont{{Matsuda}}},
  \bibinfo{author}{\bibfnamefont{U.}~\bibnamefont{{Anzer}}},
  \bibinfo{author}{\bibfnamefont{G.}~\bibnamefont{{Boerner}}},
  \bibnamefont{and} \bibinfo{author}{\bibfnamefont{M.}~\bibnamefont{{Livio}}},
  \bibinfo{journal}{Astron. and Astrophys.} \textbf{\bibinfo{volume}{221}},
  \bibinfo{pages}{263} (\bibinfo{year}{1989}).

\bibitem[{\citenamefont{{Livio} et~al.}(1991)\citenamefont{{Livio}, {Soker},
  {Matsuda}, and {Anzer}}}]{Livio1991}
\bibinfo{author}{\bibfnamefont{M.}~\bibnamefont{{Livio}}},
  \bibinfo{author}{\bibfnamefont{N.}~\bibnamefont{{Soker}}},
  \bibinfo{author}{\bibfnamefont{T.}~\bibnamefont{{Matsuda}}},
  \bibnamefont{and} \bibinfo{author}{\bibfnamefont{U.}~\bibnamefont{{Anzer}}},
  \bibinfo{journal}{Mon. Not. Roy. Astr. Soc.} \textbf{\bibinfo{volume}{253}},
  \bibinfo{pages}{633} (\bibinfo{year}{1991}).

\bibitem[{\citenamefont{{Ruffert} and {Melia}}(1994)}]{Ruffert1994}
\bibinfo{author}{\bibfnamefont{M.}~\bibnamefont{{Ruffert}}} \bibnamefont{and}
  \bibinfo{author}{\bibfnamefont{F.}~\bibnamefont{{Melia}}},
  \bibinfo{journal}{Astron. Astrophys.} \textbf{\bibinfo{volume}{288}},
  \bibinfo{pages}{L29} (\bibinfo{year}{1994}).

\bibitem[{\citenamefont{{Ruffert} and {Arnett}}(1994)}]{Ruffert1994b}
\bibinfo{author}{\bibfnamefont{M.}~\bibnamefont{{Ruffert}}} \bibnamefont{and}
  \bibinfo{author}{\bibfnamefont{D.}~\bibnamefont{{Arnett}}},
  \bibinfo{journal}{Astrophys. J.} \textbf{\bibinfo{volume}{427}},
  \bibinfo{pages}{351} (\bibinfo{year}{1994}).

\bibitem[{\citenamefont{{Ruffert}}(1997)}]{Ruffert1997}
\bibinfo{author}{\bibfnamefont{M.}~\bibnamefont{{Ruffert}}},
  \bibinfo{journal}{Astron. Astrophys.} \textbf{\bibinfo{volume}{317}},
  \bibinfo{pages}{793} (\bibinfo{year}{1997}), \eprint{arXiv:astro-ph/9605072}.

\bibitem[{\citenamefont{{Foglizzo} and {Ruffert}}(1999)}]{Foglizzo1999}
\bibinfo{author}{\bibfnamefont{T.}~\bibnamefont{{Foglizzo}}} \bibnamefont{and}
  \bibinfo{author}{\bibfnamefont{M.}~\bibnamefont{{Ruffert}}},
  \bibinfo{journal}{Astron. Astrophys.} \textbf{\bibinfo{volume}{347}},
  \bibinfo{pages}{901} (\bibinfo{year}{1999}).

\bibitem[{\citenamefont{{Foglizzo} et~al.}(2005)\citenamefont{{Foglizzo},
  {Galletti}, and {Ruffert}}}]{Foglizzo2005}
\bibinfo{author}{\bibfnamefont{T.}~\bibnamefont{{Foglizzo}}},
  \bibinfo{author}{\bibfnamefont{P.}~\bibnamefont{{Galletti}}},
  \bibnamefont{and}
  \bibinfo{author}{\bibfnamefont{M.}~\bibnamefont{{Ruffert}}},
  \bibinfo{journal}{Astron. Astrophys.} \textbf{\bibinfo{volume}{435}},
  \bibinfo{pages}{397} (\bibinfo{year}{2005}), \eprint{arXiv:astro-ph/0502168}.

\bibitem[{\citenamefont{El~Mellah and Casse}(2015)}]{Mellah:2015sja}
\bibinfo{author}{\bibfnamefont{I.}~\bibnamefont{El~Mellah}} \bibnamefont{and}
  \bibinfo{author}{\bibfnamefont{F.}~\bibnamefont{Casse}},
  \bibinfo{journal}{Mon. Not. Roy. Astron. Soc.}
  \textbf{\bibinfo{volume}{454}}, \bibinfo{pages}{2657} (\bibinfo{year}{2015}),
  \eprint{1509.07700}.

\bibitem[{\citenamefont{Beckmann et~al.}(2018)\citenamefont{Beckmann, Slyz, and
  Devriendt}}]{Beckmann:2018xfi}
\bibinfo{author}{\bibfnamefont{R.~S.} \bibnamefont{Beckmann}},
  \bibinfo{author}{\bibfnamefont{A.}~\bibnamefont{Slyz}}, \bibnamefont{and}
  \bibinfo{author}{\bibfnamefont{J.}~\bibnamefont{Devriendt}},
  \bibinfo{journal}{Mon. Not. Roy. Astron. Soc.}
  \textbf{\bibinfo{volume}{478}}, \bibinfo{pages}{995} (\bibinfo{year}{2018}),
  \eprint{1803.03014}.

\bibitem[{\citenamefont{{Edgar}}(2004)}]{Edgar2004}
\bibinfo{author}{\bibfnamefont{R.}~\bibnamefont{{Edgar}}},
  \bibinfo{journal}{New Astronomy Review} \textbf{\bibinfo{volume}{48}},
  \bibinfo{pages}{843} (\bibinfo{year}{2004}), \eprint{arXiv:astro-ph/0406166}.

\bibitem[{\citenamefont{{Rezzolla} and {Zanotti}}(2013)}]{Rezzolla_book:2013}
\bibinfo{author}{\bibfnamefont{L.}~\bibnamefont{{Rezzolla}}} \bibnamefont{and}
  \bibinfo{author}{\bibfnamefont{O.}~\bibnamefont{{Zanotti}}},
  \emph{\bibinfo{title}{Relativistic Hydrodynamics}}
  (\bibinfo{publisher}{Oxford University Press}, \bibinfo{address}{Oxford, UK},
  \bibinfo{year}{2013}), ISBN \bibinfo{isbn}{9780198528906}.

\bibitem[{\citenamefont{{Cruz-Osorio} and {Rezzolla}}(2020)}]{Cruz2020b}
\bibinfo{author}{\bibfnamefont{A.}~\bibnamefont{{Cruz-Osorio}}}
  \bibnamefont{and}
  \bibinfo{author}{\bibfnamefont{L.}~\bibnamefont{{Rezzolla}}},
  \bibinfo{journal}{Astrophys. J.} \textbf{\bibinfo{volume}{894}},
  \bibinfo{eid}{147} (\bibinfo{year}{2020}), \eprint{2004.13782}.

\bibitem[{\citenamefont{{Petrich} et~al.}(1989)\citenamefont{{Petrich},
  {Shapiro}, {Stark}, and {Teukolsky}}}]{Petrich89}
\bibinfo{author}{\bibfnamefont{L.~I.} \bibnamefont{{Petrich}}},
  \bibinfo{author}{\bibfnamefont{S.~L.} \bibnamefont{{Shapiro}}},
  \bibinfo{author}{\bibfnamefont{R.~F.} \bibnamefont{{Stark}}},
  \bibnamefont{and} \bibinfo{author}{\bibfnamefont{S.~A.}
  \bibnamefont{{Teukolsky}}}, \bibinfo{journal}{Astrophys. J.}
  \textbf{\bibinfo{volume}{336}}, \bibinfo{pages}{313} (\bibinfo{year}{1989}).

\bibitem[{\citenamefont{Font and Ib{\'a}{\~n}ez}(1998{\natexlab{a}})}]{Font98a}
\bibinfo{author}{\bibfnamefont{J.~A.} \bibnamefont{Font}} \bibnamefont{and}
  \bibinfo{author}{\bibfnamefont{J.~M.} \bibnamefont{Ib{\'a}{\~n}ez}},
  \bibinfo{journal}{Astrophys. J.} \textbf{\bibinfo{volume}{494}},
  \bibinfo{pages}{297} (\bibinfo{year}{1998}{\natexlab{a}}).

\bibitem[{\citenamefont{Font and Ib{\'a}{\~n}ez}(1998{\natexlab{b}})}]{Font98c}
\bibinfo{author}{\bibfnamefont{J.~A.} \bibnamefont{Font}} \bibnamefont{and}
  \bibinfo{author}{\bibfnamefont{J.~M.} \bibnamefont{Ib{\'a}{\~n}ez}},
  \bibinfo{journal}{Mon. Not. R. Astron. Soc.} \textbf{\bibinfo{volume}{298}},
  \bibinfo{pages}{835} (\bibinfo{year}{1998}{\natexlab{b}}).

\bibitem[{\citenamefont{Font et~al.}(1998)\citenamefont{Font, Ib{\'a}{\~n}ez,
  and Papadopoulos}}]{Font98d}
\bibinfo{author}{\bibfnamefont{J.~A.} \bibnamefont{Font}},
  \bibinfo{author}{\bibfnamefont{J.~M.} \bibnamefont{Ib{\'a}{\~n}ez}},
  \bibnamefont{and}
  \bibinfo{author}{\bibfnamefont{P.}~\bibnamefont{Papadopoulos}},
  \bibinfo{journal}{Astrophys. J.} \textbf{\bibinfo{volume}{507}},
  \bibinfo{pages}{L67} (\bibinfo{year}{1998}).

\bibitem[{\citenamefont{{Font} et~al.}(1999)\citenamefont{{Font},
  {Ib{\'a}{\~n}ez}, and {Papadopoulos}}}]{Font1999b}
\bibinfo{author}{\bibfnamefont{J.~A.} \bibnamefont{{Font}}},
  \bibinfo{author}{\bibfnamefont{J.~M.} \bibnamefont{{Ib{\'a}{\~n}ez}}},
  \bibnamefont{and}
  \bibinfo{author}{\bibfnamefont{P.}~\bibnamefont{{Papadopoulos}}},
  \bibinfo{journal}{Mon. Not. R. Astron. Soc.} \textbf{\bibinfo{volume}{305}},
  \bibinfo{pages}{920} (\bibinfo{year}{1999}), \eprint{arXiv:astro-ph/9810344}.

\bibitem[{\citenamefont{{D{\"o}nmez} et~al.}(2011)\citenamefont{{D{\"o}nmez},
  {Zanotti}, and {Rezzolla}}}]{Donmez2010}
\bibinfo{author}{\bibfnamefont{O.}~\bibnamefont{{D{\"o}nmez}}},
  \bibinfo{author}{\bibfnamefont{O.}~\bibnamefont{{Zanotti}}},
  \bibnamefont{and}
  \bibinfo{author}{\bibfnamefont{L.}~\bibnamefont{{Rezzolla}}},
  \bibinfo{journal}{Mon. Not. R. Astron. Soc.} \textbf{\bibinfo{volume}{412}},
  \bibinfo{pages}{1659} (\bibinfo{year}{2011}), \eprint{1010.1739}.

\bibitem[{\citenamefont{{Cruz-Osorio} et~al.}(2012)\citenamefont{{Cruz-Osorio},
  {Lora-Clavijo}, and {Guzm{\'a}n}}}]{Cruz2012}
\bibinfo{author}{\bibfnamefont{A.}~\bibnamefont{{Cruz-Osorio}}},
  \bibinfo{author}{\bibfnamefont{F.~D.} \bibnamefont{{Lora-Clavijo}}},
  \bibnamefont{and} \bibinfo{author}{\bibfnamefont{F.~S.}
  \bibnamefont{{Guzm{\'a}n}}}, \bibinfo{journal}{Mon. Not. R. Astron. Soc.}
  \textbf{\bibinfo{volume}{426}}, \bibinfo{pages}{732} (\bibinfo{year}{2012}),
  \eprint{1210.6588}.

\bibitem[{\citenamefont{{Lora-Clavijo} and {Guzm{\'a}n}}(2013)}]{Lora2013}
\bibinfo{author}{\bibfnamefont{F.~D.} \bibnamefont{{Lora-Clavijo}}}
  \bibnamefont{and} \bibinfo{author}{\bibfnamefont{F.~S.}
  \bibnamefont{{Guzm{\'a}n}}}, \bibinfo{journal}{Mon. Not. R. Astron. Soc.}
  \textbf{\bibinfo{volume}{429}}, \bibinfo{pages}{3144} (\bibinfo{year}{2013}),
  \eprint{1212.2139}.

\bibitem[{\citenamefont{{Cruz-Osorio} et~al.}(2013)\citenamefont{{Cruz-Osorio},
  {Lora-Clavijo}, and {Guzm{\'a}n}}}]{Cruz2013}
\bibinfo{author}{\bibfnamefont{A.}~\bibnamefont{{Cruz-Osorio}}},
  \bibinfo{author}{\bibfnamefont{F.~D.} \bibnamefont{{Lora-Clavijo}}},
  \bibnamefont{and} \bibinfo{author}{\bibfnamefont{F.~S.}
  \bibnamefont{{Guzm{\'a}n}}}, in \emph{\bibinfo{booktitle}{American Institute
  of Physics Conference Series}}, edited by
  \bibinfo{editor}{\bibfnamefont{L.~A.} \bibnamefont{{Uren\~na-L{\'o}pez}}},
  \bibinfo{editor}{\bibfnamefont{R.}~\bibnamefont{{Becerril-B{\'a}rcenas}}},
  \bibnamefont{and}
  \bibinfo{editor}{\bibfnamefont{R.}~\bibnamefont{{Linares-Romero}}}
  (\bibinfo{year}{2013}), vol. \bibinfo{volume}{1548} of
  \emph{\bibinfo{series}{American Institute of Physics Conference Series}}, pp.
  \bibinfo{pages}{323--327}, \eprint{1307.4108},
  \urlprefix\url{https://ui.adsabs.harvard.edu/abs/2013AIPC.1548..323C}.

\bibitem[{\citenamefont{{Penner}}(2011)}]{Penner2011}
\bibinfo{author}{\bibfnamefont{A.~J.} \bibnamefont{{Penner}}},
  \bibinfo{journal}{Mon. Not. R. Astron. Soc.} p. \bibinfo{pages}{490}
  (\bibinfo{year}{2011}), \eprint{1011.2976}.

\bibitem[{\citenamefont{{Kaaz} et~al.}(2022)\citenamefont{{Kaaz},
  {Murguia-Berthier}, {Chatterjee}, {Liska}, and {Tchekhovskoy}}}]{Kaaz2022}
\bibinfo{author}{\bibfnamefont{N.}~\bibnamefont{{Kaaz}}},
  \bibinfo{author}{\bibfnamefont{A.}~\bibnamefont{{Murguia-Berthier}}},
  \bibinfo{author}{\bibfnamefont{K.}~\bibnamefont{{Chatterjee}}},
  \bibinfo{author}{\bibfnamefont{M.}~\bibnamefont{{Liska}}}, \bibnamefont{and}
  \bibinfo{author}{\bibfnamefont{A.}~\bibnamefont{{Tchekhovskoy}}},
  \bibinfo{journal}{arXiv e-prints} \bibinfo{eid}{arXiv:2201.11753}
  (\bibinfo{year}{2022}), \eprint{2201.11753}.

\bibitem[{\citenamefont{{Gracia-Linares} and
  {Guzm{\'a}n}}(2023)}]{Gracia-Linares2023}
\bibinfo{author}{\bibfnamefont{M.}~\bibnamefont{{Gracia-Linares}}}
  \bibnamefont{and} \bibinfo{author}{\bibfnamefont{F.~S.}
  \bibnamefont{{Guzm{\'a}n}}}, \bibinfo{journal}{arXiv e-prints}
  \bibinfo{eid}{arXiv:2301.04307} (\bibinfo{year}{2023}), \eprint{2301.04307}.

\bibitem[{\citenamefont{{Zanotti} et~al.}(2011)\citenamefont{{Zanotti},
  {Roedig}, {Rezzolla}, and {Del Zanna}}}]{Zanotti2011}
\bibinfo{author}{\bibfnamefont{O.}~\bibnamefont{{Zanotti}}},
  \bibinfo{author}{\bibfnamefont{C.}~\bibnamefont{{Roedig}}},
  \bibinfo{author}{\bibfnamefont{L.}~\bibnamefont{{Rezzolla}}},
  \bibnamefont{and} \bibinfo{author}{\bibfnamefont{L.}~\bibnamefont{{Del
  Zanna}}}, \bibinfo{journal}{Mon. Not. R. Astron. Soc.}
  \textbf{\bibinfo{volume}{417}}, \bibinfo{pages}{2899} (\bibinfo{year}{2011}),
  \eprint{1105.5615}.

\bibitem[{\citenamefont{{Lora-Clavijo}
  et~al.}(2015{\natexlab{a}})\citenamefont{{Lora-Clavijo}, {Cruz-Osorio}, and
  {Moreno M{\'e}ndez}}}]{Lora2015219}
\bibinfo{author}{\bibfnamefont{F.~D.} \bibnamefont{{Lora-Clavijo}}},
  \bibinfo{author}{\bibfnamefont{A.}~\bibnamefont{{Cruz-Osorio}}},
  \bibnamefont{and} \bibinfo{author}{\bibfnamefont{E.}~\bibnamefont{{Moreno
  M{\'e}ndez}}}, \bibinfo{journal}{Astrophys. J., Supp.}
  \textbf{\bibinfo{volume}{219}}, \bibinfo{eid}{30}
  (\bibinfo{year}{2015}{\natexlab{a}}), \eprint{1506.08713}.

\bibitem[{\citenamefont{{Cruz-Osorio} and {Lora-Clavijo}}(2016)}]{Cruz2016}
\bibinfo{author}{\bibfnamefont{A.}~\bibnamefont{{Cruz-Osorio}}}
  \bibnamefont{and} \bibinfo{author}{\bibfnamefont{F.~D.}
  \bibnamefont{{Lora-Clavijo}}}, \bibinfo{journal}{Mon. Not. R. Astron. Soc.}
  \textbf{\bibinfo{volume}{460}}, \bibinfo{pages}{3193} (\bibinfo{year}{2016}),
  \eprint{1605.04176}.

\bibitem[{\citenamefont{Cruz-Osorio et~al.}(2017)\citenamefont{Cruz-Osorio,
  Sanchez-Salcedo, and Lora-Clavijo}}]{Cruz2017}
\bibinfo{author}{\bibfnamefont{A.}~\bibnamefont{Cruz-Osorio}},
  \bibinfo{author}{\bibfnamefont{F.~J.} \bibnamefont{Sanchez-Salcedo}},
  \bibnamefont{and} \bibinfo{author}{\bibfnamefont{F.~D.}
  \bibnamefont{Lora-Clavijo}}, \bibinfo{journal}{Mon. Not. Roy. Astron. Soc.}
  \textbf{\bibinfo{volume}{471}}, \bibinfo{pages}{3127} (\bibinfo{year}{2017}),
  \eprint{1707.05548}, \urlprefix\url{https://doi.org/10.1093/mnras/stx1815}.

\bibitem[{\citenamefont{{Read} and {Gilmore}}(2003)}]{Read2003}
\bibinfo{author}{\bibfnamefont{J.~I.} \bibnamefont{{Read}}} \bibnamefont{and}
  \bibinfo{author}{\bibfnamefont{G.}~\bibnamefont{{Gilmore}}},
  \bibinfo{journal}{Mon. Not. R. Astron. Soc.} \textbf{\bibinfo{volume}{339}},
  \bibinfo{pages}{949} (\bibinfo{year}{2003}), \eprint{astro-ph/0210658}.

\bibitem[{\citenamefont{{King} and {Pringle}}(2006)}]{King2006}
\bibinfo{author}{\bibfnamefont{A.~R.} \bibnamefont{{King}}} \bibnamefont{and}
  \bibinfo{author}{\bibfnamefont{J.~E.} \bibnamefont{{Pringle}}},
  \bibinfo{journal}{Mon. Not. R. Astron. Soc.} \textbf{\bibinfo{volume}{373}},
  \bibinfo{pages}{L90} (\bibinfo{year}{2006}), \eprint{astro-ph/0609598}.

\bibitem[{\citenamefont{{Guzm{\'a}n} and
  {Lora-Clavijo}}(2011{\natexlab{a}})}]{Guzman2011a}
\bibinfo{author}{\bibfnamefont{F.~S.} \bibnamefont{{Guzm{\'a}n}}}
  \bibnamefont{and} \bibinfo{author}{\bibfnamefont{F.~D.}
  \bibnamefont{{Lora-Clavijo}}}, \bibinfo{journal}{Mon. Not. R. Astron. Soc.}
  \textbf{\bibinfo{volume}{415}}, \bibinfo{pages}{225}
  (\bibinfo{year}{2011}{\natexlab{a}}), \eprint{1103.5497}.

\bibitem[{\citenamefont{{Guzm{\'a}n} and
  {Lora-Clavijo}}(2011{\natexlab{b}})}]{Guzman2011b}
\bibinfo{author}{\bibfnamefont{F.~S.} \bibnamefont{{Guzm{\'a}n}}}
  \bibnamefont{and} \bibinfo{author}{\bibfnamefont{F.~D.}
  \bibnamefont{{Lora-Clavijo}}}, \bibinfo{journal}{Mon. Not. R. Astron. Soc.}
  \textbf{\bibinfo{volume}{416}}, \bibinfo{pages}{3083}
  (\bibinfo{year}{2011}{\natexlab{b}}), \eprint{1106.3521}.

\bibitem[{\citenamefont{{Lora-Clavijo}
  et~al.}(2014)\citenamefont{{Lora-Clavijo}, {Gracia-Linares}, and
  {Guzm{\'a}n}}}]{Lora2014}
\bibinfo{author}{\bibfnamefont{F.~D.} \bibnamefont{{Lora-Clavijo}}},
  \bibinfo{author}{\bibfnamefont{M.}~\bibnamefont{{Gracia-Linares}}},
  \bibnamefont{and} \bibinfo{author}{\bibfnamefont{F.~S.}
  \bibnamefont{{Guzm{\'a}n}}}, \bibinfo{journal}{Mon. Not. R. Astron. Soc.}
  \textbf{\bibinfo{volume}{443}}, \bibinfo{pages}{2242} (\bibinfo{year}{2014}),
  \eprint{1406.7233},
  \urlprefix\url{https://ui.adsabs.harvard.edu/abs/2014MNRAS.443.2242L}.

\bibitem[{\citenamefont{Matos and Guzman}(2000)}]{Matos1998vk}
\bibinfo{author}{\bibfnamefont{T.}~\bibnamefont{Matos}} \bibnamefont{and}
  \bibinfo{author}{\bibfnamefont{F.~S.} \bibnamefont{Guzman}},
  \bibinfo{journal}{Class. Quant. Grav.} \textbf{\bibinfo{volume}{17}},
  \bibinfo{pages}{L9} (\bibinfo{year}{2000}), \eprint{gr-qc/9810028}.

\bibitem[{\citenamefont{Matos et~al.}(2000)\citenamefont{Matos, Guzman, and
  Nunez}}]{Matos2000}
\bibinfo{author}{\bibfnamefont{T.}~\bibnamefont{Matos}},
  \bibinfo{author}{\bibfnamefont{F.~S.} \bibnamefont{Guzman}},
  \bibnamefont{and} \bibinfo{author}{\bibfnamefont{D.}~\bibnamefont{Nunez}},
  \bibinfo{journal}{Phys. Rev.} \textbf{\bibinfo{volume}{D62}},
  \bibinfo{pages}{061301} (\bibinfo{year}{2000}), \eprint{astro-ph/0003398}.

\bibitem[{\citenamefont{Sahni and Wang}(2000)}]{Sahni:1999qe}
\bibinfo{author}{\bibfnamefont{V.}~\bibnamefont{Sahni}} \bibnamefont{and}
  \bibinfo{author}{\bibfnamefont{L.-M.} \bibnamefont{Wang}},
  \bibinfo{journal}{Phys. Rev.} \textbf{\bibinfo{volume}{D62}},
  \bibinfo{pages}{103517} (\bibinfo{year}{2000}), \eprint{astro-ph/9910097}.

\bibitem[{\citenamefont{Matos and Urena-Lopez}(2000)}]{Matos:2000ng}
\bibinfo{author}{\bibfnamefont{T.}~\bibnamefont{Matos}} \bibnamefont{and}
  \bibinfo{author}{\bibfnamefont{L.~A.} \bibnamefont{Urena-Lopez}},
  \bibinfo{journal}{Class. Quant. Grav.} \textbf{\bibinfo{volume}{17}},
  \bibinfo{pages}{L75} (\bibinfo{year}{2000}), \eprint{astro-ph/0004332}.

\bibitem[{\citenamefont{Urena-Lopez and Liddle}(2002)}]{UrenaLopez:2002du}
\bibinfo{author}{\bibfnamefont{L.~A.} \bibnamefont{Urena-Lopez}}
  \bibnamefont{and} \bibinfo{author}{\bibfnamefont{A.~R.}
  \bibnamefont{Liddle}}, \bibinfo{journal}{Phys. Rev.}
  \textbf{\bibinfo{volume}{D66}}, \bibinfo{pages}{083005}
  (\bibinfo{year}{2002}), \eprint{astro-ph/0207493}.

\bibitem[{\citenamefont{Lora-Clavijo et~al.}(2010)\citenamefont{Lora-Clavijo,
  Cruz-Osorio, and Guzman}}]{LoraClavijo:2010xc}
\bibinfo{author}{\bibfnamefont{F.~D.} \bibnamefont{Lora-Clavijo}},
  \bibinfo{author}{\bibfnamefont{A.}~\bibnamefont{Cruz-Osorio}},
  \bibnamefont{and} \bibinfo{author}{\bibfnamefont{F.~S.}
  \bibnamefont{Guzman}}, \bibinfo{journal}{Phys. Rev.}
  \textbf{\bibinfo{volume}{D82}}, \bibinfo{pages}{023005}
  (\bibinfo{year}{2010}), \eprint{1007.1162},
  \urlprefix\url{https://doi.org/10.1103/PhysRevD.82.023005}.

\bibitem[{\citenamefont{Cruz-Osorio et~al.}(2011)\citenamefont{Cruz-Osorio,
  Guzman, and Lora-Clavijo}}]{CruzOsorio:2010qs}
\bibinfo{author}{\bibfnamefont{A.}~\bibnamefont{Cruz-Osorio}},
  \bibinfo{author}{\bibfnamefont{F.~S.} \bibnamefont{Guzman}},
  \bibnamefont{and} \bibinfo{author}{\bibfnamefont{F.~D.}
  \bibnamefont{Lora-Clavijo}}, \bibinfo{journal}{Journal of Cosmology and
  Astroparticle Physics} \textbf{\bibinfo{volume}{1106}}, \bibinfo{pages}{029}
  (\bibinfo{year}{2011}), \eprint{1008.0027},
  \urlprefix\url{https://iopscience.iop.org/article/10.1088/1475-7516/2011/06/029}.

\bibitem[{\citenamefont{{Cruz-Osorio}
  et~al.}(2010{\natexlab{a}})\citenamefont{{Cruz-Osorio}, {Lora-Clavijo}, and
  {Guzm{\'a}n}}}]{Cruz2010}
\bibinfo{author}{\bibfnamefont{A.}~\bibnamefont{{Cruz-Osorio}}},
  \bibinfo{author}{\bibfnamefont{F.~D.} \bibnamefont{{Lora-Clavijo}}},
  \bibnamefont{and} \bibinfo{author}{\bibfnamefont{F.~S.}
  \bibnamefont{{Guzm{\'a}n}}}, in \emph{\bibinfo{booktitle}{American Institute
  of Physics Conference Series}}, edited by
  \bibinfo{editor}{\bibfnamefont{H.~A.} \bibnamefont{{Morales-Tecotl}}},
  \bibinfo{editor}{\bibfnamefont{L.~A.} \bibnamefont{{Urena-Lopez}}},
  \bibinfo{editor}{\bibfnamefont{R.}~\bibnamefont{{Linares-Romero}}},
  \bibnamefont{and} \bibinfo{editor}{\bibfnamefont{H.~H.}
  \bibnamefont{{Garcia-Compean}}} (\bibinfo{year}{2010}{\natexlab{a}}), vol.
  \bibinfo{volume}{1256} of \emph{\bibinfo{series}{American Institute of
  Physics Conference Series}}, pp. \bibinfo{pages}{311--317},
  \urlprefix\url{https://doi.org/10.1063/1.3473871}.

\bibitem[{\citenamefont{{Cruz-Osorio}
  et~al.}(2010{\natexlab{b}})\citenamefont{{Cruz-Osorio}, {Gonzalez-Juarez},
  {Guzman}, and {Lora-Clavijo}}}]{Cruz2010a}
\bibinfo{author}{\bibfnamefont{A.}~\bibnamefont{{Cruz-Osorio}}},
  \bibinfo{author}{\bibfnamefont{A.}~\bibnamefont{{Gonzalez-Juarez}}},
  \bibinfo{author}{\bibfnamefont{F.~S.} \bibnamefont{{Guzman}}},
  \bibnamefont{and} \bibinfo{author}{\bibfnamefont{F.~D.}
  \bibnamefont{{Lora-Clavijo}}}, \bibinfo{journal}{Rev. Mex. Fis.}
  \textbf{\bibinfo{volume}{56}}, \bibinfo{pages}{456}
  (\bibinfo{year}{2010}{\natexlab{b}}), \eprint{arXiv:1007.3776},
  \urlprefix\url{https://rmf.smf.mx/ojs/index.php/rmf/article/view/3788}.

\bibitem[{\citenamefont{{Barranco} and {Bernal}}(2011)}]{Barranco2011}
\bibinfo{author}{\bibfnamefont{J.}~\bibnamefont{{Barranco}}} \bibnamefont{and}
  \bibinfo{author}{\bibfnamefont{A.}~\bibnamefont{{Bernal}}},
  \bibinfo{journal}{Phys. Rev. D} \textbf{\bibinfo{volume}{83}},
  \bibinfo{eid}{043525} (\bibinfo{year}{2011}), \eprint{1001.1769}.

\bibitem[{\citenamefont{{Barranco} et~al.}(2012)\citenamefont{{Barranco},
  {Bernal}, {Degollado}, {Diez-Tejedor}, {Megevand}, {Alcubierre}, {N\'u\~nez},
  and {Sarbach}}}]{Barranco2012}
\bibinfo{author}{\bibfnamefont{J.}~\bibnamefont{{Barranco}}},
  \bibinfo{author}{\bibfnamefont{A.}~\bibnamefont{{Bernal}}},
  \bibinfo{author}{\bibfnamefont{J.~C.} \bibnamefont{{Degollado}}},
  \bibinfo{author}{\bibfnamefont{A.}~\bibnamefont{{Diez-Tejedor}}},
  \bibinfo{author}{\bibfnamefont{M.}~\bibnamefont{{Megevand}}},
  \bibinfo{author}{\bibfnamefont{M.}~\bibnamefont{{Alcubierre}}},
  \bibinfo{author}{\bibfnamefont{D.}~\bibnamefont{{N\'u\~nez}}},
  \bibnamefont{and}
  \bibinfo{author}{\bibfnamefont{O.}~\bibnamefont{{Sarbach}}},
  \bibinfo{journal}{Phys. Rev. Lett.} \textbf{\bibinfo{volume}{109}},
  \bibinfo{pages}{081102} (\bibinfo{year}{2012}),
  \urlprefix\url{https://link.aps.org/doi/10.1103/PhysRevLett.109.081102}.

\bibitem[{\citenamefont{Guzman and Lora-Clavijo}(2012)}]{Guzman:2012jc}
\bibinfo{author}{\bibfnamefont{F.~S.} \bibnamefont{Guzman}} \bibnamefont{and}
  \bibinfo{author}{\bibfnamefont{F.~D.} \bibnamefont{Lora-Clavijo}},
  \bibinfo{journal}{Phys. Rev.} \textbf{\bibinfo{volume}{D85}},
  \bibinfo{pages}{024036} (\bibinfo{year}{2012}), \eprint{1201.3598}.

\bibitem[{\citenamefont{{Witek} et~al.}(2013)\citenamefont{{Witek}, {Cardoso},
  {Ishibashi}, and {Sperhake}}}]{Witek2013}
\bibinfo{author}{\bibfnamefont{H.}~\bibnamefont{{Witek}}},
  \bibinfo{author}{\bibfnamefont{V.}~\bibnamefont{{Cardoso}}},
  \bibinfo{author}{\bibfnamefont{A.}~\bibnamefont{{Ishibashi}}},
  \bibnamefont{and}
  \bibinfo{author}{\bibfnamefont{U.}~\bibnamefont{{Sperhake}}},
  \bibinfo{journal}{Phys. Rev. D} \textbf{\bibinfo{volume}{87}},
  \bibinfo{eid}{043513} (\bibinfo{year}{2013}), \eprint{1212.0551}.

\bibitem[{\citenamefont{{Sanchis-Gual}
  et~al.}(2015)\citenamefont{{Sanchis-Gual}, {Degollado}, {Montero}, and
  {Font}}}]{Sanchis2015}
\bibinfo{author}{\bibfnamefont{N.}~\bibnamefont{{Sanchis-Gual}}},
  \bibinfo{author}{\bibfnamefont{J.~C.} \bibnamefont{{Degollado}}},
  \bibinfo{author}{\bibfnamefont{P.~J.} \bibnamefont{{Montero}}},
  \bibnamefont{and} \bibinfo{author}{\bibfnamefont{J.~A.}
  \bibnamefont{{Font}}}, \bibinfo{journal}{Phys. Rev. D}
  \textbf{\bibinfo{volume}{91}}, \bibinfo{eid}{043005} (\bibinfo{year}{2015}),
  \eprint{1412.8304}.

\bibitem[{\citenamefont{{Barranco} et~al.}(2017)\citenamefont{{Barranco},
  {Bernal}, {Degollado}, {Diez-Tejedor}, {Megevand}, {N\'u\~nez}, and
  {Sarbach}}}]{Barranco2017}
\bibinfo{author}{\bibfnamefont{J.}~\bibnamefont{{Barranco}}},
  \bibinfo{author}{\bibfnamefont{A.}~\bibnamefont{{Bernal}}},
  \bibinfo{author}{\bibfnamefont{J.~C.} \bibnamefont{{Degollado}}},
  \bibinfo{author}{\bibfnamefont{A.}~\bibnamefont{{Diez-Tejedor}}},
  \bibinfo{author}{\bibfnamefont{M.}~\bibnamefont{{Megevand}}},
  \bibinfo{author}{\bibfnamefont{D.}~\bibnamefont{{N\'u\~nez}}},
  \bibnamefont{and}
  \bibinfo{author}{\bibfnamefont{O.}~\bibnamefont{{Sarbach}}},
  \bibinfo{journal}{Phys. Rev. D} \textbf{\bibinfo{volume}{96}},
  \bibinfo{pages}{024049} (\bibinfo{year}{2017}),
  \urlprefix\url{https://link.aps.org/doi/10.1103/PhysRevD.96.024049}.

\bibitem[{\citenamefont{{Aguilar-Nieto}
  et~al.}(2022)\citenamefont{{Aguilar-Nieto}, {Jaramillo}, {Barranco},
  {Bernal}, {Degollado}, and {N{\'u}{\~n}ez}}}]{Aguilar2022}
\bibinfo{author}{\bibfnamefont{A.}~\bibnamefont{{Aguilar-Nieto}}},
  \bibinfo{author}{\bibfnamefont{V.}~\bibnamefont{{Jaramillo}}},
  \bibinfo{author}{\bibfnamefont{J.}~\bibnamefont{{Barranco}}},
  \bibinfo{author}{\bibfnamefont{A.}~\bibnamefont{{Bernal}}},
  \bibinfo{author}{\bibfnamefont{J.~C.} \bibnamefont{{Degollado}}},
  \bibnamefont{and}
  \bibinfo{author}{\bibfnamefont{D.}~\bibnamefont{{N{\'u}{\~n}ez}}},
  \bibinfo{journal}{arXiv e-prints} \bibinfo{eid}{arXiv:2211.10456}
  (\bibinfo{year}{2022}), \eprint{2211.10456}.

\bibitem[{\citenamefont{{Degollado} and {Herdeiro}}(2014)}]{Degollado2014}
\bibinfo{author}{\bibfnamefont{J.~C.} \bibnamefont{{Degollado}}}
  \bibnamefont{and} \bibinfo{author}{\bibfnamefont{C.~A.~R.}
  \bibnamefont{{Herdeiro}}}, \bibinfo{journal}{Phys. Rev. D}
  \textbf{\bibinfo{volume}{90}}, \bibinfo{eid}{065019} (\bibinfo{year}{2014}),
  \eprint{1408.2589}.

\bibitem[{\citenamefont{Herdeiro and Radu}(2014)}]{Herdeiro2014}
\bibinfo{author}{\bibfnamefont{C.~A.~R.} \bibnamefont{Herdeiro}}
  \bibnamefont{and} \bibinfo{author}{\bibfnamefont{E.}~\bibnamefont{Radu}},
  \bibinfo{journal}{Phys. Rev. Lett.} \textbf{\bibinfo{volume}{112}},
  \bibinfo{pages}{221101} (\bibinfo{year}{2014}),
  \urlprefix\url{https://link.aps.org/doi/10.1103/PhysRevLett.112.221101}.

\bibitem[{\citenamefont{{Herdeiro} and {Radu}}(2015)}]{Herdeiro2015}
\bibinfo{author}{\bibfnamefont{C.}~\bibnamefont{{Herdeiro}}} \bibnamefont{and}
  \bibinfo{author}{\bibfnamefont{E.}~\bibnamefont{{Radu}}},
  \bibinfo{journal}{Classical and Quantum Gravity}
  \textbf{\bibinfo{volume}{32}}, \bibinfo{eid}{144001} (\bibinfo{year}{2015}).

\bibitem[{\citenamefont{{Ganchev} and {Santos}}(2018)}]{Ganchev2018}
\bibinfo{author}{\bibfnamefont{B.}~\bibnamefont{{Ganchev}}} \bibnamefont{and}
  \bibinfo{author}{\bibfnamefont{J.~E.} \bibnamefont{{Santos}}},
  \bibinfo{journal}{Phys. Rev. Lett.} \textbf{\bibinfo{volume}{120}},
  \bibinfo{eid}{171101} (\bibinfo{year}{2018}), \eprint{1711.08464}.

\bibitem[{\citenamefont{{Degollado} et~al.}(2018)\citenamefont{{Degollado},
  {Herdeiro}, and {Radu}}}]{Degollado2018}
\bibinfo{author}{\bibfnamefont{J.~C.} \bibnamefont{{Degollado}}},
  \bibinfo{author}{\bibfnamefont{C.~A.~R.} \bibnamefont{{Herdeiro}}},
  \bibnamefont{and} \bibinfo{author}{\bibfnamefont{E.}~\bibnamefont{{Radu}}},
  \bibinfo{journal}{Physics Letters B} \textbf{\bibinfo{volume}{781}},
  \bibinfo{pages}{651} (\bibinfo{year}{2018}), \eprint{1802.07266}.

\bibitem[{\citenamefont{Cunha et~al.}(2015)\citenamefont{Cunha, Herdeiro, Radu,
  and R\'unarsson}}]{Cunha2015}
\bibinfo{author}{\bibfnamefont{P.~V.~P.} \bibnamefont{Cunha}},
  \bibinfo{author}{\bibfnamefont{C.~A.~R.} \bibnamefont{Herdeiro}},
  \bibinfo{author}{\bibfnamefont{E.}~\bibnamefont{Radu}}, \bibnamefont{and}
  \bibinfo{author}{\bibfnamefont{H.~F.} \bibnamefont{R\'unarsson}},
  \bibinfo{journal}{Phys. Rev. Lett.} \textbf{\bibinfo{volume}{115}},
  \bibinfo{pages}{211102} (\bibinfo{year}{2015}), \eprint{1509.00021},
  \urlprefix\url{https://link.aps.org/doi/10.1103/PhysRevLett.115.211102}.

\bibitem[{\citenamefont{Cunha et~al.}(2016)\citenamefont{Cunha, Grover,
  Herdeiro, Radu, Runarsson, and Wittig}}]{Cunha:2016}
\bibinfo{author}{\bibfnamefont{P.}~\bibnamefont{Cunha}},
  \bibinfo{author}{\bibfnamefont{J.}~\bibnamefont{Grover}},
  \bibinfo{author}{\bibfnamefont{C.}~\bibnamefont{Herdeiro}},
  \bibinfo{author}{\bibfnamefont{E.}~\bibnamefont{Radu}},
  \bibinfo{author}{\bibfnamefont{H.}~\bibnamefont{Runarsson}},
  \bibnamefont{and} \bibinfo{author}{\bibfnamefont{A.}~\bibnamefont{Wittig}},
  \bibinfo{journal}{Phys. Rev. D} \textbf{\bibinfo{volume}{94}},
  \bibinfo{pages}{104023} (\bibinfo{year}{2016}), \eprint{1609.01340}.

\bibitem[{\citenamefont{Cunha et~al.}(2019)\citenamefont{Cunha, Herdeiro, and
  Radu}}]{Cunha:2019ikd}
\bibinfo{author}{\bibfnamefont{P.~V.} \bibnamefont{Cunha}},
  \bibinfo{author}{\bibfnamefont{C.~A.} \bibnamefont{Herdeiro}},
  \bibnamefont{and} \bibinfo{author}{\bibfnamefont{E.}~\bibnamefont{Radu}},
  \bibinfo{journal}{Universe} \textbf{\bibinfo{volume}{5}},
  \bibinfo{pages}{220} (\bibinfo{year}{2019}), \eprint{1909.08039}.

\bibitem[{\citenamefont{{Delgado} et~al.}(2016)\citenamefont{{Delgado},
  {Herdeiro}, {Radu}, and {R{\'u}narsson}}}]{Delgado2016}
\bibinfo{author}{\bibfnamefont{J.~F.~M.} \bibnamefont{{Delgado}}},
  \bibinfo{author}{\bibfnamefont{C.~A.~R.} \bibnamefont{{Herdeiro}}},
  \bibinfo{author}{\bibfnamefont{E.}~\bibnamefont{{Radu}}}, \bibnamefont{and}
  \bibinfo{author}{\bibfnamefont{H.}~\bibnamefont{{R{\'u}narsson}}},
  \bibinfo{journal}{Physics Letters B} \textbf{\bibinfo{volume}{761}},
  \bibinfo{pages}{234} (\bibinfo{year}{2016}), \eprint{1608.00631}.

\bibitem[{\citenamefont{Franchini et~al.}(2017)\citenamefont{Franchini, Pani,
  Maselli, Gualtieri, Herdeiro, Radu, and Ferrari}}]{Franchini2016}
\bibinfo{author}{\bibfnamefont{N.}~\bibnamefont{Franchini}},
  \bibinfo{author}{\bibfnamefont{P.}~\bibnamefont{Pani}},
  \bibinfo{author}{\bibfnamefont{A.}~\bibnamefont{Maselli}},
  \bibinfo{author}{\bibfnamefont{L.}~\bibnamefont{Gualtieri}},
  \bibinfo{author}{\bibfnamefont{C.~A.~R.} \bibnamefont{Herdeiro}},
  \bibinfo{author}{\bibfnamefont{E.}~\bibnamefont{Radu}}, \bibnamefont{and}
  \bibinfo{author}{\bibfnamefont{V.}~\bibnamefont{Ferrari}},
  \bibinfo{journal}{Phys. Rev.} \textbf{\bibinfo{volume}{D95}},
  \bibinfo{pages}{124025} (\bibinfo{year}{2017}), \eprint{1612.00038}.

\bibitem[{\citenamefont{Ni et~al.}(2016)\citenamefont{Ni, Zhou,
  Cardenas-Avendano, Bambi, Herdeiro, and Radu}}]{Ni2016a}
\bibinfo{author}{\bibfnamefont{Y.}~\bibnamefont{Ni}},
  \bibinfo{author}{\bibfnamefont{M.}~\bibnamefont{Zhou}},
  \bibinfo{author}{\bibfnamefont{A.}~\bibnamefont{Cardenas-Avendano}},
  \bibinfo{author}{\bibfnamefont{C.}~\bibnamefont{Bambi}},
  \bibinfo{author}{\bibfnamefont{C.~A.~R.} \bibnamefont{Herdeiro}},
  \bibnamefont{and} \bibinfo{author}{\bibfnamefont{E.}~\bibnamefont{Radu}},
  \bibinfo{journal}{Journal of Cosmology and Astroparticle Physics}
  \textbf{\bibinfo{volume}{07}}, \bibinfo{pages}{049} (\bibinfo{year}{2016}),
  \eprint{1606.04654}.

\bibitem[{\citenamefont{{Gimeno-Soler}
  et~al.}(2019)\citenamefont{{Gimeno-Soler}, {Font}, {Herdeiro}, and
  {Radu}}}]{Gimeno-Soler:2019}
\bibinfo{author}{\bibfnamefont{S.}~\bibnamefont{{Gimeno-Soler}}},
  \bibinfo{author}{\bibfnamefont{J.~A.} \bibnamefont{{Font}}},
  \bibinfo{author}{\bibfnamefont{C.}~\bibnamefont{{Herdeiro}}},
  \bibnamefont{and} \bibinfo{author}{\bibfnamefont{E.}~\bibnamefont{{Radu}}},
  \bibinfo{journal}{Phys. Rev. D} \textbf{\bibinfo{volume}{99}},
  \bibinfo{eid}{043002} (\bibinfo{year}{2019}), \eprint{1811.11492}.

\bibitem[{\citenamefont{Gimeno-Soler et~al.}(2021)\citenamefont{Gimeno-Soler,
  Font, Herdeiro, and Radu}}]{Gimeno-Soler2021}
\bibinfo{author}{\bibfnamefont{S.}~\bibnamefont{Gimeno-Soler}},
  \bibinfo{author}{\bibfnamefont{J.~A.} \bibnamefont{Font}},
  \bibinfo{author}{\bibfnamefont{C.}~\bibnamefont{Herdeiro}}, \bibnamefont{and}
  \bibinfo{author}{\bibfnamefont{E.}~\bibnamefont{Radu}},
  \bibinfo{journal}{Phys. Rev. D} \textbf{\bibinfo{volume}{104}},
  \bibinfo{pages}{103008} (\bibinfo{year}{2021}), \eprint{2106.15425}.

\bibitem[{\citenamefont{Banyuls et~al.}(1997)\citenamefont{Banyuls, Font,
  Ib{\'a}{\~n}ez, Mart{\'\i}, and Miralles}}]{Banyuls97}
\bibinfo{author}{\bibfnamefont{F.}~\bibnamefont{Banyuls}},
  \bibinfo{author}{\bibfnamefont{J.~A.} \bibnamefont{Font}},
  \bibinfo{author}{\bibfnamefont{J.~M.} \bibnamefont{Ib{\'a}{\~n}ez}},
  \bibinfo{author}{\bibfnamefont{J.~M.} \bibnamefont{Mart{\'\i}}},
  \bibnamefont{and} \bibinfo{author}{\bibfnamefont{J.~A.}
  \bibnamefont{Miralles}}, \bibinfo{journal}{Astrophys. J.}
  \textbf{\bibinfo{volume}{476}}, \bibinfo{pages}{221} (\bibinfo{year}{1997}).

\bibitem[{\citenamefont{{Porth} et~al.}(2017)\citenamefont{{Porth}, {Olivares},
  {Mizuno}, {Younsi}, {Rezzolla}, {Moscibrodzka}, {Falcke}, and
  {Kramer}}}]{Porth2017}
\bibinfo{author}{\bibfnamefont{O.}~\bibnamefont{{Porth}}},
  \bibinfo{author}{\bibfnamefont{H.}~\bibnamefont{{Olivares}}},
  \bibinfo{author}{\bibfnamefont{Y.}~\bibnamefont{{Mizuno}}},
  \bibinfo{author}{\bibfnamefont{Z.}~\bibnamefont{{Younsi}}},
  \bibinfo{author}{\bibfnamefont{L.}~\bibnamefont{{Rezzolla}}},
  \bibinfo{author}{\bibfnamefont{M.}~\bibnamefont{{Moscibrodzka}}},
  \bibinfo{author}{\bibfnamefont{H.}~\bibnamefont{{Falcke}}}, \bibnamefont{and}
  \bibinfo{author}{\bibfnamefont{M.}~\bibnamefont{{Kramer}}},
  \bibinfo{journal}{Computational Astrophysics and Cosmology}
  \textbf{\bibinfo{volume}{4}}, \bibinfo{eid}{1} (\bibinfo{year}{2017}),
  \eprint{1611.09720}.

\bibitem[{\citenamefont{{Olivares S{\'{a}}nchez}
  et~al.}(2018)\citenamefont{{Olivares S{\'{a}}nchez}, {Porth}, and
  {Mizuno}}}]{Olivares2018a}
\bibinfo{author}{\bibfnamefont{H.}~\bibnamefont{{Olivares S{\'{a}}nchez}}},
  \bibinfo{author}{\bibfnamefont{O.}~\bibnamefont{{Porth}}}, \bibnamefont{and}
  \bibinfo{author}{\bibfnamefont{Y.}~\bibnamefont{{Mizuno}}},
  \bibinfo{journal}{J. Phys. Conf. Ser.} \textbf{\bibinfo{volume}{1031}},
  \bibinfo{pages}{012008} (\bibinfo{year}{2018}), ISSN
  \bibinfo{issn}{1742-6588},
  \urlprefix\url{http://stacks.iop.org/1742-6596/1031/i=1/a=012008?key=crossref.f3a1d22bc66dbcfa7943b00d4b6047f9}.

\bibitem[{\citenamefont{Olivares et~al.}(2019)\citenamefont{Olivares, Porth,
  Davelaar, Most, Fromm, Mizuno, Younsi, and Rezzolla}}]{Olivares2019}
\bibinfo{author}{\bibfnamefont{H.}~\bibnamefont{Olivares}},
  \bibinfo{author}{\bibfnamefont{O.}~\bibnamefont{Porth}},
  \bibinfo{author}{\bibfnamefont{J.}~\bibnamefont{Davelaar}},
  \bibinfo{author}{\bibfnamefont{E.~R.} \bibnamefont{Most}},
  \bibinfo{author}{\bibfnamefont{C.~M.} \bibnamefont{Fromm}},
  \bibinfo{author}{\bibfnamefont{Y.}~\bibnamefont{Mizuno}},
  \bibinfo{author}{\bibfnamefont{Z.}~\bibnamefont{Younsi}}, \bibnamefont{and}
  \bibinfo{author}{\bibfnamefont{L.}~\bibnamefont{Rezzolla}},
  \bibinfo{journal}{Astron. Astrophys.} \textbf{\bibinfo{volume}{629}},
  \bibinfo{pages}{A61} (\bibinfo{year}{2019}), ISSN \bibinfo{issn}{0004-6361},
  \urlprefix\url{https://www.aanda.org/10.1051/0004-6361/201935559}.

\bibitem[{\citenamefont{{Lora-Clavijo}
  et~al.}(2015{\natexlab{b}})\citenamefont{{Lora-Clavijo}, {Cruz-Osorio}, and
  {Guzm{\'a}n}}}]{Lora2015}
\bibinfo{author}{\bibfnamefont{F.~D.} \bibnamefont{{Lora-Clavijo}}},
  \bibinfo{author}{\bibfnamefont{A.}~\bibnamefont{{Cruz-Osorio}}},
  \bibnamefont{and} \bibinfo{author}{\bibfnamefont{F.~S.}
  \bibnamefont{{Guzm{\'a}n}}}, \bibinfo{journal}{Astrophys. J., Supp.}
  \textbf{\bibinfo{volume}{218}}, \bibinfo{eid}{24}
  (\bibinfo{year}{2015}{\natexlab{b}}), \eprint{1408.5846}.

\bibitem[{\citenamefont{{L\"{o}hner}}(1987)}]{Loehner87}
\bibinfo{author}{\bibfnamefont{R.}~\bibnamefont{{L\"{o}hner}}},
  \bibinfo{journal}{Computer Methods in Applied Mechanics and Engineering}
  \textbf{\bibinfo{volume}{61}}, \bibinfo{pages}{323} (\bibinfo{year}{1987}).

\bibitem[{\citenamefont{Harten et~al.}(1983)\citenamefont{Harten, Lax, and van
  Leer}}]{Harten83}
\bibinfo{author}{\bibfnamefont{A.}~\bibnamefont{Harten}},
  \bibinfo{author}{\bibfnamefont{P.~D.} \bibnamefont{Lax}}, \bibnamefont{and}
  \bibinfo{author}{\bibfnamefont{B.}~\bibnamefont{van Leer}},
  \bibinfo{journal}{SIAM Rev.} \textbf{\bibinfo{volume}{25}},
  \bibinfo{pages}{35} (\bibinfo{year}{1983}).

\bibitem[{\citenamefont{Einfeldt}(1988)}]{Einfeldt88}
\bibinfo{author}{\bibfnamefont{B.}~\bibnamefont{Einfeldt}},
  \bibinfo{journal}{SIAM J. Numer. Anal.} \textbf{\bibinfo{volume}{25}},
  \bibinfo{pages}{294} (\bibinfo{year}{1988}).

\bibitem[{\citenamefont{{Abbott} et~al.}(2019)\citenamefont{{Abbott}, {Abbott},
  {Abbott}, {Abraham}, {Acernese}, {Ackley}, {Adams}, {Adhikari}, {Adya},
  {Affeldt} et~al.}}]{Abbott2019c}
\bibinfo{author}{\bibfnamefont{B.~P.} \bibnamefont{{Abbott}}},
  \bibinfo{author}{\bibfnamefont{R.}~\bibnamefont{{Abbott}}},
  \bibinfo{author}{\bibfnamefont{T.~D.} \bibnamefont{{Abbott}}},
  \bibinfo{author}{\bibfnamefont{S.}~\bibnamefont{{Abraham}}},
  \bibinfo{author}{\bibfnamefont{F.}~\bibnamefont{{Acernese}}},
  \bibinfo{author}{\bibfnamefont{K.}~\bibnamefont{{Ackley}}},
  \bibinfo{author}{\bibfnamefont{C.}~\bibnamefont{{Adams}}},
  \bibinfo{author}{\bibfnamefont{R.~X.} \bibnamefont{{Adhikari}}},
  \bibinfo{author}{\bibfnamefont{V.~B.} \bibnamefont{{Adya}}},
  \bibinfo{author}{\bibfnamefont{C.}~\bibnamefont{{Affeldt}}},
  \bibnamefont{et~al.}, \bibinfo{journal}{Physical Review X}
  \textbf{\bibinfo{volume}{9}}, \bibinfo{eid}{031040} (\bibinfo{year}{2019}),
  \eprint{1811.12907}.

\bibitem[{\citenamefont{{Abbott} et~al.}(2021)\citenamefont{{Abbott}, {Abbott},
  {Abraham}, {Acernese}, {Ackley}, {Adams}, {Adams}, {Adhikari}, {Adya},
  {Affeldt} et~al.}}]{Abbott2021c}
\bibinfo{author}{\bibfnamefont{R.}~\bibnamefont{{Abbott}}},
  \bibinfo{author}{\bibfnamefont{T.~D.} \bibnamefont{{Abbott}}},
  \bibinfo{author}{\bibfnamefont{S.}~\bibnamefont{{Abraham}}},
  \bibinfo{author}{\bibfnamefont{F.}~\bibnamefont{{Acernese}}},
  \bibinfo{author}{\bibfnamefont{K.}~\bibnamefont{{Ackley}}},
  \bibinfo{author}{\bibfnamefont{A.}~\bibnamefont{{Adams}}},
  \bibinfo{author}{\bibfnamefont{C.}~\bibnamefont{{Adams}}},
  \bibinfo{author}{\bibfnamefont{R.~X.} \bibnamefont{{Adhikari}}},
  \bibinfo{author}{\bibfnamefont{V.~B.} \bibnamefont{{Adya}}},
  \bibinfo{author}{\bibfnamefont{C.}~\bibnamefont{{Affeldt}}},
  \bibnamefont{et~al.}, \bibinfo{journal}{Physical Review X}
  \textbf{\bibinfo{volume}{11}}, \bibinfo{eid}{021053} (\bibinfo{year}{2021}),
  \eprint{2010.14527}.

\bibitem[{\citenamefont{{The LIGO Scientific Collaboration}
  et~al.}(2021)\citenamefont{{The LIGO Scientific Collaboration}, {the Virgo
  Collaboration}, {the KAGRA Collaboration}, {Abbott}, {Abbott}, {Acernese},
  {Ackley}, {Adams}, {Adhikari}, {Adhikari} et~al.}}]{LVK2021}
\bibinfo{author}{\bibnamefont{{The LIGO Scientific Collaboration}}},
  \bibinfo{author}{\bibnamefont{{the Virgo Collaboration}}},
  \bibinfo{author}{\bibnamefont{{the KAGRA Collaboration}}},
  \bibinfo{author}{\bibfnamefont{R.}~\bibnamefont{{Abbott}}},
  \bibinfo{author}{\bibfnamefont{T.~D.} \bibnamefont{{Abbott}}},
  \bibinfo{author}{\bibfnamefont{F.}~\bibnamefont{{Acernese}}},
  \bibinfo{author}{\bibfnamefont{K.}~\bibnamefont{{Ackley}}},
  \bibinfo{author}{\bibfnamefont{C.}~\bibnamefont{{Adams}}},
  \bibinfo{author}{\bibfnamefont{N.}~\bibnamefont{{Adhikari}}},
  \bibinfo{author}{\bibfnamefont{R.~X.} \bibnamefont{{Adhikari}}},
  \bibnamefont{et~al.}, \bibinfo{journal}{arXiv e-prints}
  \bibinfo{eid}{arXiv:2111.03606} (\bibinfo{year}{2021}), \eprint{2111.03606}.

\bibitem[{\citenamefont{{De Luca} et~al.}(2021{\natexlab{b}})\citenamefont{{De
  Luca}, {Franciolini}, {Pani}, and {Riotto}}}]{DeLuca2021b}
\bibinfo{author}{\bibfnamefont{V.}~\bibnamefont{{De Luca}}},
  \bibinfo{author}{\bibfnamefont{G.}~\bibnamefont{{Franciolini}}},
  \bibinfo{author}{\bibfnamefont{P.}~\bibnamefont{{Pani}}}, \bibnamefont{and}
  \bibinfo{author}{\bibfnamefont{A.}~\bibnamefont{{Riotto}}},
  \bibinfo{journal}{Journal of Cosmology and Astroparticle Physics}
  \textbf{\bibinfo{volume}{2021}}, \bibinfo{eid}{003}
  (\bibinfo{year}{2021}{\natexlab{b}}), \eprint{2102.03809}.

\bibitem[{\citenamefont{{Rezzolla}
  et~al.}(2003{\natexlab{a}})\citenamefont{{Rezzolla}, {Yoshida}, {Maccarone},
  and {Zanotti}}}]{Rezzolla_qpo_03a}
\bibinfo{author}{\bibfnamefont{L.}~\bibnamefont{{Rezzolla}}},
  \bibinfo{author}{\bibfnamefont{S.}~\bibnamefont{{Yoshida}}},
  \bibinfo{author}{\bibfnamefont{T.~J.} \bibnamefont{{Maccarone}}},
  \bibnamefont{and}
  \bibinfo{author}{\bibfnamefont{O.}~\bibnamefont{{Zanotti}}},
  \bibinfo{journal}{Mon. Not. R. Astron. Soc.} \textbf{\bibinfo{volume}{344}},
  \bibinfo{pages}{L37} (\bibinfo{year}{2003}{\natexlab{a}}),
  \eprint{arXiv:astro-ph/0307487}.

\bibitem[{\citenamefont{{Rezzolla}
  et~al.}(2003{\natexlab{b}})\citenamefont{{Rezzolla}, {Yoshida}, and
  {Zanotti}}}]{Rezzolla_qpo_03b}
\bibinfo{author}{\bibfnamefont{L.}~\bibnamefont{{Rezzolla}}},
  \bibinfo{author}{\bibfnamefont{S.}~\bibnamefont{{Yoshida}}},
  \bibnamefont{and}
  \bibinfo{author}{\bibfnamefont{O.}~\bibnamefont{{Zanotti}}},
  \bibinfo{journal}{Mon. Not. R. Astron. Soc.} \textbf{\bibinfo{volume}{344}},
  \bibinfo{pages}{978} (\bibinfo{year}{2003}{\natexlab{b}}),
  \eprint{arXiv:astro-ph/0307488}.

\bibitem[{\citenamefont{{Zanotti} et~al.}(2005)\citenamefont{{Zanotti}, {Font},
  {Rezzolla}, and {Montero}}}]{Zanotti05}
\bibinfo{author}{\bibfnamefont{O.}~\bibnamefont{{Zanotti}}},
  \bibinfo{author}{\bibfnamefont{J.~A.} \bibnamefont{{Font}}},
  \bibinfo{author}{\bibfnamefont{L.}~\bibnamefont{{Rezzolla}}},
  \bibnamefont{and} \bibinfo{author}{\bibfnamefont{P.~J.}
  \bibnamefont{{Montero}}}, \bibinfo{journal}{Mon. Not. R. Astron. Soc.}
  \textbf{\bibinfo{volume}{356}}, \bibinfo{pages}{1371} (\bibinfo{year}{2005}),
  \eprint{arXiv:astro-ph/0411116}.

\bibitem[{\citenamefont{{T{\"o}r{\"o}k}}(2005)}]{Torok2005}
\bibinfo{author}{\bibfnamefont{G.}~\bibnamefont{{T{\"o}r{\"o}k}}},
  \bibinfo{journal}{Astron. Astrophys.} \textbf{\bibinfo{volume}{440}},
  \bibinfo{pages}{1} (\bibinfo{year}{2005}), \eprint{arXiv:astro-ph/0412500}.

\bibitem[{\citenamefont{{Schnittman} and {Rezzolla}}(2006)}]{Schnittman06}
\bibinfo{author}{\bibfnamefont{J.~D.} \bibnamefont{{Schnittman}}}
  \bibnamefont{and}
  \bibinfo{author}{\bibfnamefont{L.}~\bibnamefont{{Rezzolla}}},
  \bibinfo{journal}{Astrophys. J.} \textbf{\bibinfo{volume}{637}},
  \bibinfo{pages}{L113} (\bibinfo{year}{2006}),
  \eprint{arXiv:astro-ph/0506702}.

\bibitem[{\citenamefont{{Orosz} et~al.}(2002)\citenamefont{{Orosz}, {Groot},
  {van der Klis}, {McClintock}, {Garcia}, {Zhao}, {Jain}, {Bailyn}, and
  {Remillard}}}]{Orosz2002}
\bibinfo{author}{\bibfnamefont{J.~A.} \bibnamefont{{Orosz}}},
  \bibinfo{author}{\bibfnamefont{P.~J.} \bibnamefont{{Groot}}},
  \bibinfo{author}{\bibfnamefont{M.}~\bibnamefont{{van der Klis}}},
  \bibinfo{author}{\bibfnamefont{J.~E.} \bibnamefont{{McClintock}}},
  \bibinfo{author}{\bibfnamefont{M.~R.} \bibnamefont{{Garcia}}},
  \bibinfo{author}{\bibfnamefont{P.}~\bibnamefont{{Zhao}}},
  \bibinfo{author}{\bibfnamefont{R.~K.} \bibnamefont{{Jain}}},
  \bibinfo{author}{\bibfnamefont{C.~D.} \bibnamefont{{Bailyn}}},
  \bibnamefont{and} \bibinfo{author}{\bibfnamefont{R.~A.}
  \bibnamefont{{Remillard}}}, \bibinfo{journal}{Astrophys. J.}
  \textbf{\bibinfo{volume}{568}}, \bibinfo{pages}{845} (\bibinfo{year}{2002}),
  \eprint{astro-ph/0112101}.

\end{thebibliography}

\end{document}